\documentclass[12pt]{article}

\usepackage{amsmath,amssymb,mathrsfs,cite,bm,color}
\usepackage{graphicx}

\newcommand{\Tr}{\mathrm{Tr}}

\newcommand{\beq}{\begin{equation}}
\newcommand{\eeq}{\end{equation}}

\makeatletter
 
  \@addtoreset{equation}{section}
 \makeatother

\allowdisplaybreaks

\begin{document}
\begin{titlepage}
\renewcommand{\thefootnote}{\fnsymbol{footnote}}
\begin{flushright}
DIAS-STP-22-17\\
UTHEP-776\\
YITP-22-132
\end{flushright}
\vspace{12mm}
\begin{center}
{\Large 
On the existence of the NS5-brane limit of\\
the plane wave matrix model
}
\end{center}
\vspace{7mm}
\begin{center}
Yuhma~A{\sc sano}$^{\phantom{(}1), 2)}$\footnote{
e-mail address: 
asano@het.ph.tsukuba.ac.jp
}, 
Goro~I{\sc shiki}$^{\phantom{(}1), 2)}$\footnote{
e-mail address: 
ishiki@het.ph.tsukuba.ac.jp
}, 
Takaki~M{\sc atsumoto}$^{\phantom{(}3)}$\footnote{
e-mail address: 
takaki@stp.dias.ie
},\\
Shinji~S{\sc himasaki}\footnote{
e-mail address: 
shimasaki.s@gmail.com
}
    {\sc and}
Hiromasa~W{\sc atanabe}$^{\phantom{(}4)}$\footnote{
e-mail address: 
hiromasa.watanabe@yukawa.kyoto-u.ac.jp
}

\par \vspace{7mm}

$^{1)}$ {\it 
Graduate School of Pure and Applied Sciences, University of Tsukuba,\\
Tsukuba, Ibaraki 305-8571, Japan
}\\

$^{2)}$ {\it 
Tomonaga Center for the History of the Universe, University of Tsukuba,\\
Tsukuba, Ibaraki 305-8571, Japan
}\\

$^{3)}$ {\it 
School of Theoretical Physics, Dublin Institute for Advanced Studies,\\
10 Burlington Road, Dublin 4, Ireland
}\\

$^{4)}$ {\it 
Yukawa Institute for Theoretical Physics, Kyoto University,\\
Kyoto 606-8502, Japan
}\\

% $^{4)}$ {\it 
% Research and Education Center for Natural Sciences, Keio University, 
% Hiyoshi 4-1-1, Yokohama, Kanagawa 223-8521, Japan
% }\\

\end{center}
% \vspace{7mm}
\newpage
\begin{abstract}\noindent
We consider a double scaling limit of the plane wave matrix model
(PWMM), 
in which the gravity dual geometry of PWMM reduces to a class of 
spherical NS5-brane solutions.
We identify the form of the scaling limit for 
the dual geometry of PWMM around a general vacuum
and then translate the limit into the field theoretic 
language. 
We also show 
that the limit indeed exists at least in a certain planar
1/4-BPS sector of PWMM
by using the localization computation analytically.
In addition, we employ the hybrid Monte Carlo method 
to compute the matrix integral obtained by the localization method,
near the parameter region where the supergravity approximation is valid.
Our numerical results,
which are considered to be the first computation of 
quantum loop correction to the Lin-Maldacena geometry,
suggest that the double scaling limit exists
beyond the planar sector.

\end{abstract}
\setcounter{footnote}{0}
\end{titlepage}

\tableofcontents

%%%%%%%%%%%%%%%%%%%%%%%%%%%%%%%%%%%%%%%%%%%%%%%%%%%%%%%%%%%%%
\section{Introduction}
%%%%%%%%%%%%%%%%%%%%%%%%%%%%%%%%%%%%%%%%%%%%%%%%%%%%%%%%%%%%%
The gauge/gravity correspondence claims the equivalence between 
a certain class of gauge theories and string theories
\cite{Maldacena, GKP, Witten}. 
The dual description provided by this relation often gives a 
useful method of analyzing the gauge theories and string theories. 
This is also applicable to the little string theory (LST), which 
is considered to be defined on NS5-branes in the decoupling limit
\cite{LSTreview}.
Although any direct definition of LST using a Lagrangian
has not been established yet, the gravity description enables us to 
extract information of LST.

%The little string theory (LST) is defined as 
%a theory living on NS5-branes in the decoupling limit
%and is considered to have an interesting feature that 
%it has both stringy and non-gravitational field theoretic properties.
%Though the the semi-classical bulk gravity description provides 
%a method of analyzing this theory, 
%there is no known direct definition of the theory at this moment.
%In order to clarify the nature of fivebranes, it will be important 
%to give more direct description of LST.

% Recently, 
Based on the gauge/gravity duality,
it was conjectured that the type IIA LST on $R\times S^5$ is described 
by the plane wave matrix model (PWMM) \cite{Berenstein:2002jq} 
in a double scaling limit \cite{LM, Ling:2006up},
which we call ``the NS5-brane limit'' in this paper.
The gravity dual of PWMM possesses a scaling limit where 
the geometry reduces to an NS5-brane solution. 
If the gauge/gravity correspondence is true, 
it implies that the limit also exists on the (gauge-)field theory side.
Thus, the IIA LST is expected to be described as this limit of PWMM.
Since PWMM is a well-defined matrix quantum mechanics, 
this relation is expected to give a direct definition of the LST.
In this paper, we investigate this relation.

%between $SU(2|4)$ symmetric gauge theories \cite{LLM,LM}, 
%There are some proposals for such direct description.
%Among those, we focus on the proposal made in \cite{Ling:2006up} for
%type IIA LST on $R\times S^5$.
%it was conjectured that the type IIA LST on $R\times S^5$ is described 
%by the plane wave matrix model (PWMM) in a double scaling limit 

The gauge/gravity duality in this context
is for theories with $SU(2|4)$ symmetry.
The field theory side in this duality consists of
% The theories considered in this context 
% have $SU(2|4)$ symmetry and consist of four theories: 
PWMM, $\mathcal{N}=8$ SYM on $R\times S^2$, 
$\mathcal{N}=4$ SYM on $R\times S^3/Z_k$ and the IIA LST on $R\times S^5$
\cite{LM} (see also \cite{LLM,Ishiki:2006yr,Ishiki:2006rt}).
These theories have many discrete vacua, which preserve the $SU(2|4)$ symmetry.
The vacua in PWMM are given by fuzzy spheres and classified 
by the representation of the $SU(2)$ Lie algebra. 
For SYM on $R\times S^2$ and SYM on $R\times S^3/Z_k$,
the vacua are labeled by monopole charges and holonomy, respectively.
The description of the vacua of the IIA LST is not clear due to
the lack of the direct definition of the theory.
However, from the gravity description, one can find that 
each non-trivial vacuum of the LST carries not only the NS5-brane charges but 
also some D2-brane charges\footnote{These branes are compact, 
so that the net brane charge must be vanishing in the Matrix theory description
\cite{Maldacena:2002rb}.
We mean by ``the brane charges'' the local dipole charges in this paper.}.

For each vacuum of these theories, the $SU(2|4)$ symmetric dual geometry 
was constructed \cite{LM}.
This class of geometries has $R\times SO(3) \times SO(6)$ isometry, 
and they locally contain $R\times S^2\times S^5$.
A remarkable feature %of those geometries 
is that the equations that determine those geometries 
can be rewritten into the form of the Laplace equations in certain 
axially symmetric electrostatic systems.
The system consists of background potential and 
a set of conducting disks,
% The electrostatic systems consist of some conducting disks and 
% the corresponding gauge theory is specified
% by the boundary condition for the axial direction.
% The geometries are thus labeled by the disk configurations 
% as well as the boundary conditions.
% The background potential and the disk configuration
which specify the corresponding field theory
and a vacuum of the theory, respectively.
The geometry is completely fixed by the electrostatic potential
of the system.

The geometry dual to the IIA LST is a gravity solution for spherical NS5-branes.
It was pointed out that this NS5-brane geometry can be obtained by taking 
a limit of the dual geometry of PWMM such that one of the conducting
disks becomes infinitely large \cite{LM,Ling:2006up}. 
Then, this limit is translated into the language on the field theory side, 
and found to be a certain double scaling limit in PWMM, in which the 
't~Hooft coupling also becomes large as the matrix size approaches infinity.
Through the gauge/gravity correspondence, this relation suggests that
the IIA LST is realized as the double scaling limit of PWMM.
The same argument can also be made for $\mathcal{N}=8$ SYM on $R\times S^2$ and
$\mathcal{N}=4$ SYM on $R\times S^3/Z_k$ \cite{Ling:2006xi}. 
%, namely, 
%the IIA LST can also be described as some double scaling limits of 
%these theories .

The form of the double scaling limit was found only for 
the case of a simple vacuum of PWMM \cite{Ling:2006up}.
In this case, the resulting IIA LST is the theory around 
the trivial vacuum of the LST, which has only one stack of fivebranes.
In this paper, we generalize this argument to the situations
where the IIA LST around a general non-trivial vacuum are realized, 
which also carries some D2-brane charges.
We find the form of the double scaling limit in general gravity solutions and 
then rewrite it using the field theoretic parameters.

%In \cite{Ling:2006up,Ling:2006xi}, the forms of the 
%scaling limits were found by analyzing the gravity solutions. 
% and then 
%the limits are translated into the language of the 
%field theoretic parameters based on some natural arguments.
For this purpose, we use the remarkable result found 
in \cite{Asano:2014vba,Asano:2014eca} based 
on the localization computation\footnote{
Beside this result \cite{Asano:2014vba,Asano:2014eca},
there are some applications of the localisation computation
\cite{Asano:2017xiy,Asano:2017nxw,Roychowdhury:2021unp}. 
} 
which states that the eigenvalue distribution of an adjoint scalar field
on the gauge theory side is identified with the charge densities on 
the conducting disks.
Through this relation, 
we can identify the parameters on both sides and 
express the double scaling limit in terms of the parameters of PWMM.
%The form of the double scaling limit we find in this paper turns out to be
%a natural generalization of the limit obtained
%in \cite{Ling:2006up,Ling:2006xi}.
Furthermore, the direct identification between the eigenvalue
densities and the charge densities also 
% partially shows 
makes it possible to see that the double scaling limit 
exists at least in the planar quarter-BPS sector of PWMM which 
is considered in the localization computation.

In addition to the analytic computation,
we numerically compute quantities of the BPS sector in PWMM with finite matrix size $N$
by applying the hybrid Monte Carlo method to the matrix integral obtained
in the localization computation.
Such numerical computation allows us to discuss the double scaling limit 
beyond the planar sector 
because it can provide information of $1/N$ corrections.
As a result of the numerical computation, 
we find evidence that the double scaling limit seems to exist 
at the non-planar level.

This paper is organized as follows.
In section 2, 
we review how the dual geometry is related to
the electrostatic system.
In section 3, we consider the double scaling limit, in which the 
dual geometry of PWMM is reduced to an NS5-brane solution.
We first review the simplest case obtained in \cite{Ling:2006up}
and then generalize it to the cases for non-trivial vacua.
In section 4, after we review 
the correspondence between the scalar eigenvalue densities 
and the charge densities, 
we see how the double scaling limit is realized on the gauge theory side.
In section 5, 
we present numerical results that support the existence of the double scaling limit.
Finally, we summarize the paper in section 6 with discussion.

%%%%%%%%%%%%%%%%%%%%%%%%%%%%%%%%%%%%%%%%%%%%%%%%%%%%%%%%%%%%%
\section{Dual geometries of PWMM and LST} \label{dual geometries of PWMM and LST}
%%%%%%%%%%%%%%%%%%%%%%%%%%%%%%%%%%%%%%%%%%%%%%%%%%%%%%%%%%%%%
% The gravity dual geometries of the gauge theories with $SU(2|4)$
The supergravity solutions dual to the (gauge) theories with $SU(2|4)$
symmetry were obtained in \cite{LLM,LM} by using
an $SU(2|4)$ symmetric ansatz. 
% The solutions are labeled by some discrete moduli parameters 
% given by the D2- and NS5-brane charges as well as one continuous parameter, 
% which corresponds to the gauge coupling on the gauge theory side.
% The moduli parameters specify the corresponding vacuum in PWMM 
% as we describe in section \ref{NS5-brane limit in PWMM}.
The geometry is written by a single function $V(r,z)$, as
\begin{align}
 ds_{10}^2 &= 
 \left( \frac{\ddot{V}-2\dot{V}}{-V''} \right)^{1/2}
 \left\{-4 \frac{\ddot{V}}{\ddot{V}-2\dot{V}}dt^2
 -2 \frac{V''}{\dot{V}}(dr^2 +dz^2)
 +4 d\Omega_{5}^2 +2 \frac{V'' \dot{V}}{\Delta} d\Omega_2^2
 \right\}, \nonumber\\
 C_1 &= - \frac{(\dot{V}^2)'}{\ddot{V}-2\dot{V}} dt, \quad
 C_3 = -4 \frac{\dot{V}^2 V''}{\Delta} dt \wedge d\Omega_2 , 
 \nonumber\\
 B_2 &= \left( 
 \frac{(\dot{V}^2)'}{\Delta}+2z \right) d\Omega_2, \quad
 e^{4\Phi} = \frac{4(\ddot{V}-2\dot{V})^3}{-V'' \dot{V}^2 \Delta^2},
 \label{LM solution}
\end{align}
where $\Delta = (\ddot{V}-2\dot{V})V''-(\dot{V}')^2$ and 
the dots and primes indicate $\frac{\partial}{\partial \log r}$
and $\frac{\partial}{\partial z}$, respectively.
Note that the geometry contains $S^2\times S^5$,
which reflects the bosonic subgroup $SO(3)\times SO(6)$ of $SU(2|4)$.

The supersymmetry requires $V(r,z)$ to satisfy
\begin{align}
 \frac{1}{r^2}\ddot V+V''=0.
 \label{axi-Laplace-eq}
\end{align}
The positivity and regularity of the metric impose further conditions.
The positivity restricts the asymptotic behavior of $V(r,z)$ for 
large $r$ and $z$. 
The regularity
at the points where $S^2$ or $S^5$ shrinks to zero %at $\dot V=0$
requires either one of the following conditions to be satisfied:
(A) $\frac{\partial}{\partial r} V(r,z) \sim r$ at $r=0$
or (B) $z={\rm constant}$ if $\frac{\partial}{\partial r}V=0$.
One can then construct a 3-cycle or 6-cycle
% ending on the subspace given by $\dot V=0$.
that is a fiber bundle
over a non-contractible curve connecting
two points where $S^2$ or $S^5$ shrinks,
with fiber given by the sphere.
The integrals of fluxes $H_3$ and $\star \tilde F_4$ 
over those cycles yield
NS5-brane charges % $N_5^{(s)}$ 
and D2-brane charges, % $N_2^{(s)}$
respectively \cite{LM}.

The equation \eqref{axi-Laplace-eq}, which determines $V(r,z)$, is just
the Laplace equation in the axisymmetric three-dimensional space and 
the regularity condition on $V(r,z)$ can be rephrased as 
the presence of some conducting disks.
Thus, the geometry is described in terms of the 
axisymmetric electrostatic system with the conducting disks, 
and $V(r,z)$ is interpreted as an electrostatic potential.
The positivity of the metric requires 
existence of background potential in the electrostatic system, 
which we will introduce explicitly in the following subsections.

The electric charges and the $z$-coordinates of the conducting disks are parameters
of this geometry. 
As we will describe in the next subsection,
they are proportional to D2-brane and NS5-brane charges, respectively,
through the cycle integrals, 
% so that these quantities are quantized. 
and thus quantized.
On the other hand, the radii of the disks are not free parameters of the 
solution, since 
the regularity of the solution requires that the charge density on 
any finite disk must vanish at the edge.
This relates the radius of the disk to the charge.
%The parameters are related to moduli parameters in the dual gauge theories.

\subsection{Dual geometry of PWMM}
\label{Dual geometry of PWMM}
\begin{figure}
\centering
\includegraphics[width=6cm]{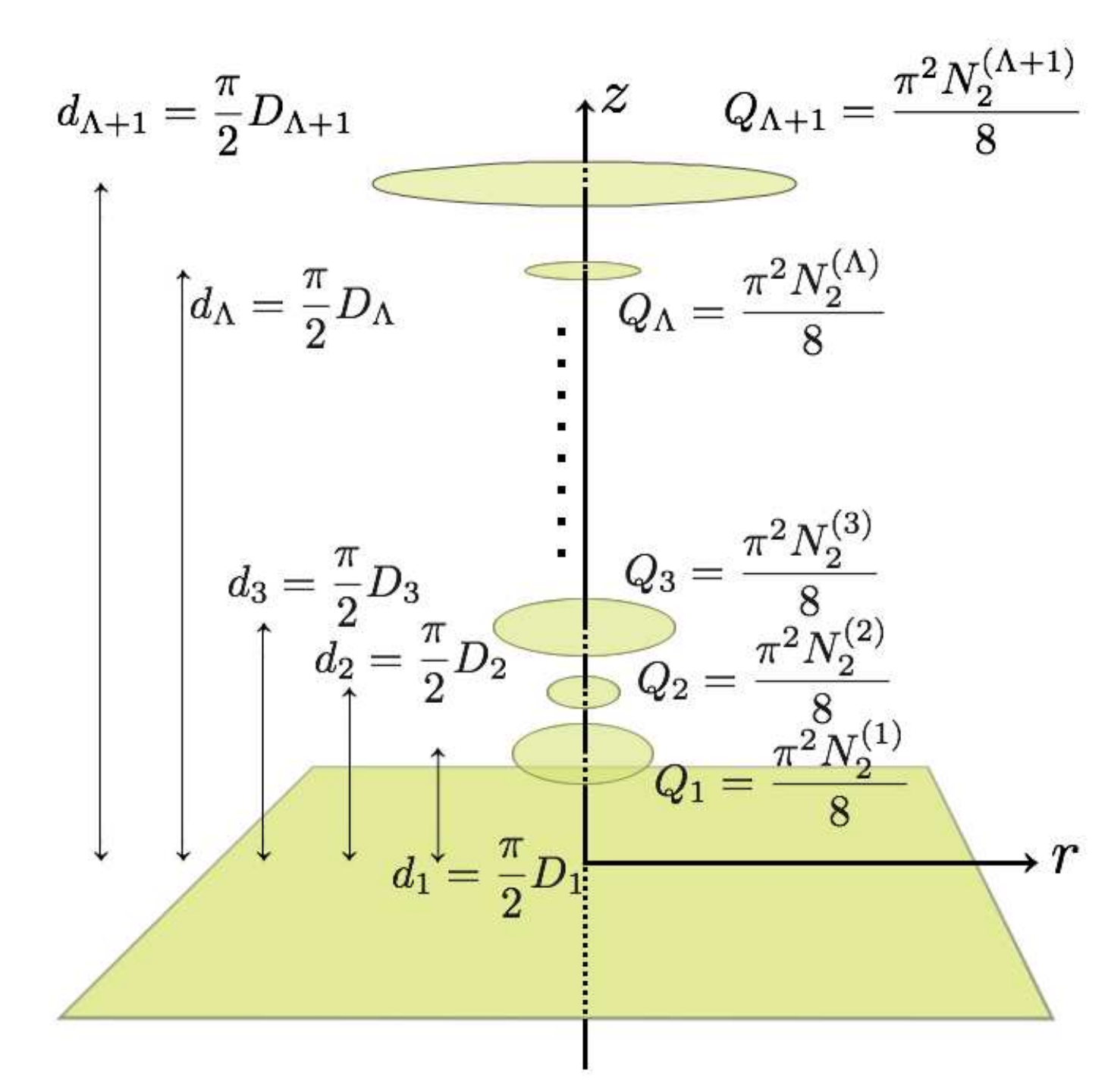}
\caption{\small The electrostatic system for a dual geometry of PWMM.}
\label{pwmm fig}
\end{figure}

In the following, we focus on the dual geometry of PWMM. 
In order for the solution to become the D0-brane geometry 
in the UV region, the presence of an infinite conducting plate (say at $z=0$) is required\footnote{When the solution is lifted up to 11 dimensions,
the infinite conducting plate is required in order for the asymptotic geometry to be the plane wave geometry.
% it is required that there exists an infinite conducting plate at $z=0$.
}.
Then only the space with $z>0$ becomes relevant for this geometry.
We also need background potential of the form
% $V(r,z)\sim V_0(r^2z-\frac{2}{3}z^3)$,
\begin{align}
 V_{\mathrm{bg}}(r,z)&=V_0\left( r^2z-\frac{2}{3}z^3 \right) ,
 \label{PWMMpot_bg}
\end{align}
where $V_0$ is a positive constant.
This is determined from the positivity of the metric. 
There are also some finite conducting disks along the $z$-direction.
We denote the $z$-coordinate and the total electric charge 
of the $s$th disk 
by $d_s$ and $Q_s$, respectively (Fig.~\ref{pwmm fig}).
Through the cycle integrals mentioned above, 
these quantities are related to % the brane charges 
the NS5-brane charge of the $s$th stack $N_5^{(s)}$ 
and the D2-brane charge of the $s$th stack $N_2^{(s)}$
as
\begin{align}
 d_s = \frac{\pi D_s}{2}, \;\;\; 
 Q_s = \frac{\pi^2 N_2^{(s)}}{8},
 \label{brane charge}
\end{align}
where $D_s=\sum_{t=1}^{s}N_5^{(t)}$.

The solution to the Laplace equation \eqref{axi-Laplace-eq}
generally takes the form \cite{vanAnders:2007ky,Asano:2014eca},
\begin{align}
 V(r,z)
 =V_{\mathrm{bg}}(r,z)
 +\sum_{s=1}^{\Lambda+1}
 V_s(r,z),
 \label{PWMMpotential}
\end{align}
where 
\begin{align}
 V_s(r,z)&=
 \frac{1}{\pi}\int_{-R_s}^{R_s}du\, 
 \left(\frac{1}{\sqrt{(z-d_s+iu)^2+r^2}}-\frac{1}{\sqrt{(z+d_s+iu)^2+r^2}}\right) f_s(u).\label{PWMMpot_s}
\end{align}
% Here, $V_{\mathrm{bg}}$ is the background potential, which ensures the asymptotic behavior.
$f_s$ and $R_s$ are an electric charge density and a radius
of the $s$th disk, respectively.
The support of the function $f_s$ is the interval $[-R_s,R_s]$. 
Let $\sigma_s (r)$ be the charge density along the radial direction
on the $s$th disk. Then, it is related to
$f_s$ by
\begin{align}
 f_s(u)=2\pi \int^{\infty}_{|u|} dr \frac{r\sigma_s(r)}{\sqrt{r^2 -u^2}},
 \qquad
 \sigma_s (r)
 =-\frac{1}{\pi^2}\int^{R_s}_r du \frac{f_s' (u)}{\sqrt{u^2 -r^2}}.
\label{fandsigma}
\end{align}
These relations imply that $f_s$ can be interpreted as 
the charge density projected onto 
% a straight line on the disk which passes through the origin of the disk.
a diameter of the disk.
The regularity of the gravity solution requires that each charge density $\sigma_s(r)$ vanish 
at the tip of the disk or equivalently $\frac{d}{du}f_s(u)$ be finite at the edges.

In terms of these expressions, 
one can show that $V_s$ satisfies the following differential equation:
\begin{align}
 \bigtriangleup V_s(r,z) 
 =-4\pi \sigma_s(r)(\delta(z-d_s)-\delta(z+d_s)).
 \label{eom_for_pot}
\end{align}
%This suggests that 
%$\sigma_s$ essentially has the same information as $V_s$ through 
%this equation.
The total charge of the $s$th disk can be written as
\begin{align}
 Q_s=2\pi \int_0^{R_s} rdr\, \sigma_s (r)
 =\frac{1}{\pi}\int_{-R_s}^{R_s} du\, f_s (u).
\label{QsRs}
\end{align}
The Laplace equation \eqref{axi-Laplace-eq} is rewritten in terms of the  
density $f_s(u)$ into the following integral equation,
\begin{align}
 f_s(u)-\sum_{t=1}^{\Lambda +1}(K_t(d_s+d_t)-K_t(|d_s-d_t|))f_t(u)
 =C_s+\frac{2}{3}V_0d_s^3-2V_0d_su^2,
 \label{PWMM saddle pt eq}
\end{align}
where $C_s$ is the constant value of the potential on the $s$th disk 
and $K_s$ is defined by
\begin{align}
 K_s(\delta)f(u):=\frac{1}{\pi}\int_{-R_s}^{R_s}du'\frac{\delta}{\delta^2+(u-u')^2}f(u'),
\label{Kf}
\end{align}
where $\delta$ is an arbitrary real number.
See \cite{vanAnders:2007ky,Asano:2014vba,Asano:2014eca} for a derivation of the 
equation \eqref{PWMM saddle pt eq}.
For general disk configurations, 
no analytic solution of (\ref{PWMM saddle pt eq}) is known. 
However, in some scaling limits considered in \cite{Ling:2006up}, 
one can evaluate the charge density.

\subsection{Dual geometry of LST}
\label{Dual geometry of LST}
\begin{figure}
\centering
\includegraphics[width=6cm]{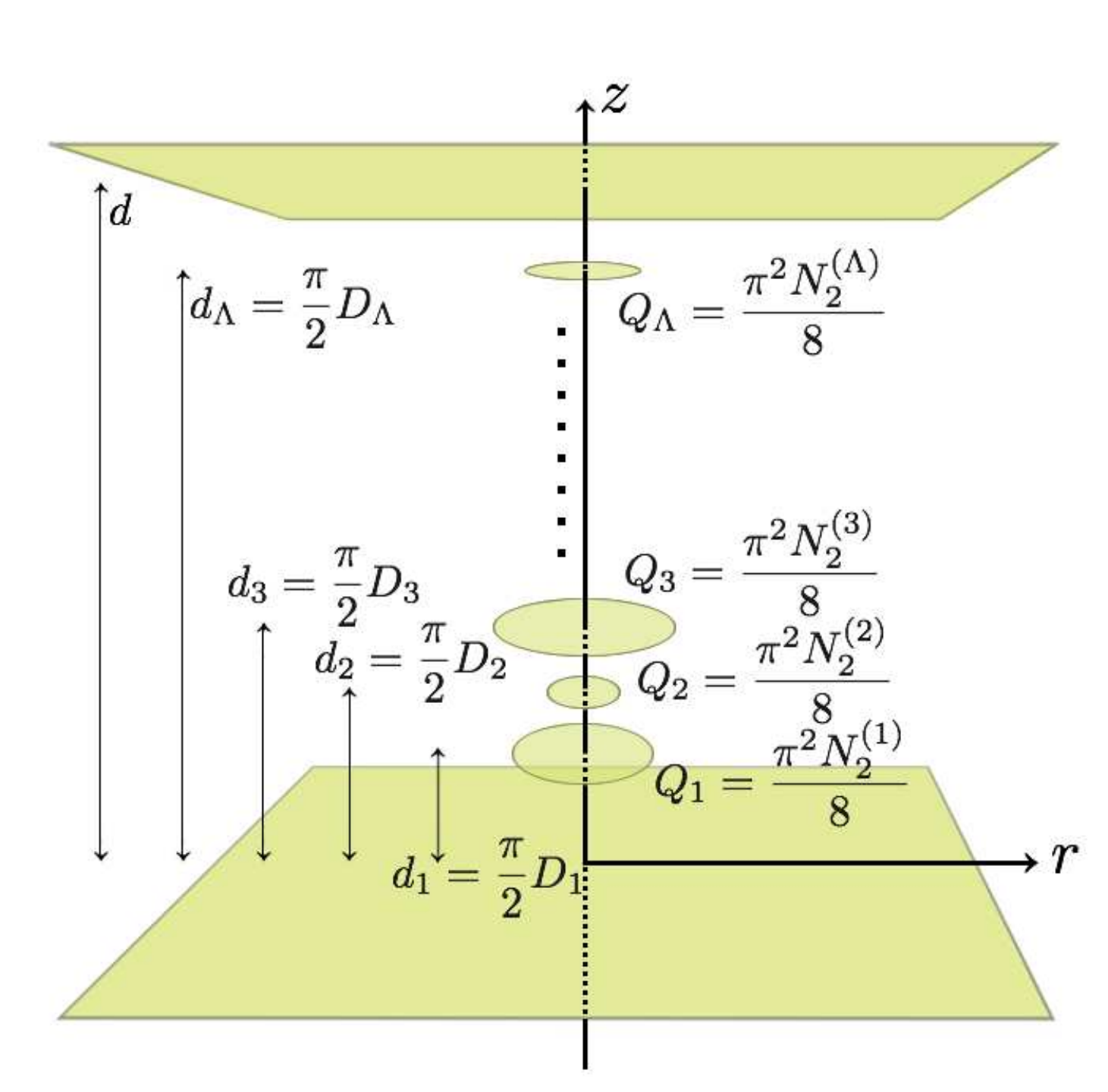}
\caption{\small The electrostatic system for a dual geometry of LST.}
\label{LST fig}
\end{figure}

The electrostatic system for a general vacuum of LST 
consists of two infinite conducting plates at $z=0$ and $z=d$
and some finite conducting disks placed between them.
The positivity of the metric requires the presence of the background potential,
\begin{align}
  \tilde V_{\mathrm{bg}}(r,z)=
 \frac{1}{g_0}\sin\frac{\pi z}{d}I_0(\frac{\pi r}{d}),
 \label{simplest NS5potential}
\end{align}
where $I_0$ is the modified Bessel function of the 0th order,
$g_0$ is a constant and $0\leq z\leq d$.
The positions and charges of the disks are related to the brane charges 
in the same way as \eqref{brane charge} (see Fig.~\ref{LST fig}).
%The form  is determined from the positivity 
%so that the $r$-$z$ space should be restricted to a periodic one 
%along the $z$-direction.

The potential for this system can be written as
\begin{align}
 \tilde V(r,z)
 &= \tilde V_{\mathrm{bg}}(r,z)
%\frac{1}{g_0}\sin\frac{\pi z}{d}I_0(\frac{\pi r}{d})
 +\sum_{n=-\infty}^{\infty}\sum_{s=1}^{\Lambda}
 \tilde V_{s,n},
 \label{NS5potential}
\end{align}
where %$0\leq z\leq d$, 
$\Lambda$ is the number of the finite conducting disks and
\begin{align}
 \tilde V_{s,n}
 =\frac{1}{\pi}\int_{-R_s}^{R_s}du\, 
 \left(\frac{1}{\sqrt{(z-d_s+2nd+iu)^2+r^2}}-\frac{1}{\sqrt{(z+d_s+2nd+iu)^2+r^2}}\right) \tilde f_s(u).
\end{align}
$\tilde f_s$ represent the charge densities on the finite conducting disks.
The infinite summation in the second term in \eqref{NS5potential} 
corresponds to contributions from infinitely many mirror images
of the conducting disks. % at $0< z< d$.
%due to the infinite conducting plates at $z=0$ and $d$.
The charge densities  $\tilde f_s$ satisfy
\begin{align}
 &\tilde f_s(u)-\sum_{n=-\infty}^{\infty}\sum_{t=1}^{\Lambda}(K_t(|d_s+d_t+2nd|)-K_t(|d_s-d_t+2nd|))\tilde f_t(u)
 \nonumber \\
 &=\tilde C_s-\frac{1}{g_0}\sin\frac{\pi d_s}{d}\cosh\frac{\pi u}{d},
 \label{tildefiIntEq}
\end{align}
where 
$\tilde{C}_s$ is the value of the potential on the $s$th disk and
$K_s(d)$ represents the integral operator defined in \eqref{Kf}.

If we consider the simplest case where there is no conducting disk between 
the two infinite plates, the geometry at large $r$ reduces to the 
geometry with linear dilaton and $H$ flux, which is 
a typical feature of NS5-brane solutions \cite{Ling:2006up}. 
This fact enables us to associate this class of solutions to the NS5-branes. 
Note that for a generic disk configuration, the geometry 
possesses not only NS5-brane charges but also D2-brane charges.

%%%%%%%%%%%%%%%%%%%%%%%%%%%%%%%%%%%%%%%%%%%%%%%%%%%%%%%%%%%%%
\section{NS5-brane limit on the gravity side}
\label{section NS5-brane limit}
%%%%%%%%%%%%%%%%%%%%%%%%%%%%%%%%%%%%%%%%%%%%%%%%%%%%%%%%%%%%%
In this section, we consider the NS5-brane limit,
in which 
the dual geometry of PWMM reviewed in section \ref{Dual geometry of PWMM}
is reduced to 
the NS5-brane geometry in section \ref{Dual geometry of LST}
as a result of a double scaling limit.

As reviewed in the previous section, the dual geometries of both PWMM and LST can be described 
in terms of electrostatic potentials of axisymmetric electrostatic systems 
(Fig.~\ref{pwmm fig} and Fig.~\ref{LST fig}).
%% The electrostatic system for a dual geometry of PWMM consists of
%% some conducting disks and one infinite conducting plate 
%% placed below them in the background potential $V\sim r^2z-\frac{2}{3}z^3$.
%% On the other hand,
%% the electrostatic system for LST consists of some conducting disks with finite radii 
%% between two infinite conducting plates in the background potential \eqref{simplest NS5potential}.
%%  where
%% the number and the size of finite conducting disks correspond to a vacuum in LST, %side,
%% while the NS5-brane and the D2-brane charges, respectively, in the gravity side.
%% %which is supposed to have many descreate vacua.
%So, the limit we should take corresponds to the limit such that
%the elecetrostatic system for PWMM turns into that for LST.
Comparing two systems, 
one can find that the NS5-brane limit corresponds to 
the limit where the radius of the disk at the highest position in $z$
in the electrostatic system for PWMM is sent to infinity.
It turns out, however, that only taking this limit does not lead to 
the electrostatic system for LST.
Indeed, in order to obtain the background potential of LST 
\eqref{simplest NS5potential},
one also needs to tune the magnitude of the background potential 
in a suitable manner.
Thus, the NS5-brane limit turns out to be a double scaling limit. 

The NS5-brane limit for the trivial vacuum of LST 
(i.e.~the case without any finite conducting disks)
was first investigated in \cite{Ling:2006up}.
Here, we extend this study to the case for a general vacuum.

\subsection{NS5-brane limit for the trivial vacuum}
%%%%%%%%%%%%%%%%%%%%%%%%%%%%%%%%%%%%%%%%%%%%%%%%%%%%%%%%%%%%%
Let us first review the NS5-brane limit for the trivial vacuum of 
LST \cite{Ling:2006up}.
%Some relations obtained in this case will be used in the general case.
The electrostatic system in this case 
consists of two infinite conducting plates
and the background potential \eqref{simplest NS5potential} (Fig.~\ref{LST fig} with $\Lambda=0$).
Since there are no charged disks, the potential of this system 
%\eqref{NS5potential}
is given only by the background potential 
(i.e.~the first term in \eqref{NS5potential}).
We will show that this electrostatic system can be obtained 
as a double scaling limit of the particular electrostatic system for PWMM 
which consists of a single finite conducting disk and the infinite plate
(Fig.~\ref{pwmm fig} with $\Lambda=0$).

In the electrostatic system of PWMM,
let $Q$, $R$ and $d$ be the total charge, 
the radius and the $z$-coordinate of the conducting disk, respectively.  
The potential of this system
is given by \eqref{PWMMpotential} and \eqref{PWMMpot_s} as
\begin{align}
 V(r,z)
 =V_0(r^2z-\frac{2}{3}z^3)+\frac{1}{\pi}\int_{-R}^{R}du\left(
 \frac{1}{\sqrt{(z-d+iu)^2+r^2}}-\frac{1}{\sqrt{(z+d+iu)^2+r^2}}
 \right) f^{(0)}(u),
 \label{simplest pp-wave potential}
\end{align}
where $0\leq z< \infty$. The radius $R$ 
is determined from the total charge $Q$ by \eqref{QsRs}:
\begin{align}
 Q=\frac{1}{\pi}\int_{-R}^{R}du\, f^{(0)}(u). \label{QR}
\end{align}
$f^{(0)}(u)$ is the charge density on the disk
% (i.e.~$f_1(u)$ in \eqref{PWMMpot_s} in this case)
and satisfies
\begin{align}
 f^{(0)}(u)
 -\frac{1}{\pi}\int_{-R}^{R}du'\frac{2d}{(2d)^2+(u-u')^2}f^{(0)}(u')
 % -K_{\Lambda+1}(2d)f^{(0)}(u)
 =C^{(0)}+\frac{2}{3}V_0d^3-2V_0d u^2,
\label{f0 eq}
\end{align}
where $C^{(0)}$ is the value of the potential \eqref{simplest pp-wave potential} on the disk at $z=d$.

Let us now find the form of the NS5-brane limit.
%by focusing on the potential in $r<R$ (the region between two disks) 
%with $R/d\gg 1$.
First, note that the boundary condition at $z=d$ of 
\eqref{simplest pp-wave potential}
does not agree with that of \eqref{simplest NS5potential}:
the value of the former is $C^{(0)}$, while the latter vanishes.
To resolve this discrepancy, we add $-C^{(0)}z/d$ to 
\eqref{simplest pp-wave potential}.
This addition of the linear term in $z$ is justified because
it does not change the supergravity solution \eqref{LM solution}.
In other words, $V(r,z)-C^{(0)}z/d$ satisfies all the requirements
for the electrostatic potential for PWMM: it
is an axisymmetric solution to the Laplace equation and 
vanishes at $z=0$.
Then, the shifted potential can be expanded in Fourier series as
\begin{align}
 V(r,z)-\frac{C^{(0)}}{d}z
 =\sum_{n=1}^{\infty}c_nI_0\left( \frac{\pi nr}{d}\right) \sin\frac{\pi nz}{d},
\label{Fourier expansion}
\end{align}
where $c_n$ are Fourier coefficients.
Note that the potential 
\eqref{simplest NS5potential} 
of the electrostatic system for LST 
appears as the lowest Fourier mode in the expansion \eqref{Fourier expansion}.
Hence, the NS5-brane limit is given by the limit 
under which only $c_1$ survives while the other $c_n$'s vanish.
From \eqref{Fourier expansion} with $r=R$, one can evaluate $c_n$ as
\begin{align}
 c_n=\left( I_0\left(\frac{\pi nR}{d}\right)\right)^{-1}\frac{2}{d}\int_0^{d}dz\,
 \left\{ V(R,z)-\frac{C^{(0)}}{d}z \right\} \sin \frac{\pi nz}{d}.
\label{cn}
\end{align}
% In the limit of $R \rightarrow \infty$,  
The leading part of the integral in (\ref{cn})
for large $R$
was evaluated in \cite{Ling:2006up} and 
found to be proportional to $V_0Rd^3$.
%Note that the order of $R$ is lower than $R^2$, 
%which is a naive estimation without taking into account any cancellations.
Then, by noting that $\left( I_0(z)\right)^{-1}\sim \sqrt{2\pi z}e^{-z}$ when $z\rightarrow \infty$,
one finds that $c_n$ behaves as $c_n \sim V_0(Rd)^{3/2}e^{-n \pi R/d}$ 
when $R\to \infty$.
Therefore, we finally obtain the NS5-brane limit as
\begin{align}
 R\to \infty,\quad V_0\to \infty
 \quad \text{with} \quad
 \frac{e^{\pi R/d}}{V_0(Rd)^{3/2}}\text{ : fixed},
 \label{simplest DSL}
\end{align}
where the fixed quantity is proportional to $g_0$ in \eqref{simplest NS5potential}.
%$V_0$ should be scaled as $V_0\sim R^{-3/2}e^{\pi R/d}$,

For later use,
let us express the relationship \eqref{QR} between $Q$ and $R$ more explicitly.
In the limit $R\to \infty$, one can solve the integral equation \eqref{f0 eq} as
\begin{align}
 f^{(0)}(u)=\frac{V_0R^3}{3}\left\{ 1-\left(\frac{u}{R}\right)^2\right\}^{\frac{3}{2}}.
\end{align}
The derivation is given in appendix \ref{apx:solution}.
%% Then, it is followed by
%% \begin{align}
%%  C^{(0)}=V_0\left( R^2d-\frac{2}{3}d^3\right) 
%%  \to V_0R^2d.
%% \end{align}
Substituting this into \eqref{QR}, one obtains
\begin{align}
 Q=\frac{V_0R^4}{8}.
\end{align}
This relation shows that the electric charge $Q$
goes to infinity in the NS5-brane limit.
Since $Q$ is proportional to the D2-brane charge by \eqref{brane charge}, 
the D2-brane charge also goes to infinity in this limit.

%%%%%%%%%%%%%%%%%%%%%%%%%%%%%%%%%%%%%%%%%%%%%%%%%%%%%%%%%%%%%
\subsection{NS5-brane limit for general vacua} \label{NS5 general}
%%%%%%%%%%%%%%%%%%%%%%%%%%%%%%%%%%%%%%%%%%%%%%%%%%%%%%%%%%%%%

Next, let us investigate the NS5-brane limit for
a general vacuum of LST. 
%the electrostatic system of which contains $\Lambda$ conducting disks between the infinite conducting plates.
The electrostatic system for the general vacuum of LST consists of $\Lambda$ 
conducting disks between
two infinite conducting plates in the presence of the background potential 
\eqref{simplest NS5potential}
(see Fig.~\ref{LST fig}).
The net electrostatic potential is given by \eqref{NS5potential}.
%The net potential of this system is given by \eqref{NS5potential}.
We will show that this electrostatic system can be obtained as 
a double scaling limit of that for PWMM 
with $\Lambda+1$ conducting disks (Fig.~\ref{pwmm fig}).
%, which we discussed in subsection 2.1,
We will find that the limit is given by
\begin{align}
 R_{\Lambda+1}\to \infty, \quad V_0\to \infty \quad \text{with}
 \quad 
% d_{\Lambda+1}=d\text{ : fixed}, 
% \quad
 \frac{e^{\pi R_{\Lambda+1}/d_{\Lambda+1}}}{V_0(R_{\Lambda+1}d_{\Lambda+1})^{3/2}}\text{ : fixed},
 % V_0\to \frac{e^{\pi R/d}}{Cg_0(Rd)^{3/2}},
 \label{DSLgrav}
\end{align}
where $R_{\Lambda+1}$ and $d_{\Lambda+1}$ are the radius and the $z$-coordinate of the highest disk, respectively.
This limit is a natural generalization of
the NS5-brane limit \eqref{simplest DSL} in the previous subsection.
%which leads to the dual geometry of the trivial vacuum of LST.
Hereafter, to simplify the notation, we denote the radius, the $z$-coordinate and the charge of the highest disk
just by $R$, $d$ and $Q$, respectively, 
instead of $R_{\Lambda+1}$, $d_{\Lambda+1}$ and $Q_{\Lambda+1}$.

%%%%%%%%%%%%%%%%%%%%%%%%%%%%%%%%%%%%%%%%%%%%%%%%%%%%%%%%%%%%%
%\subsubsection{Mirror images}
%%%%%%%%%%%%%%%%%%%%%%%%%%%%%%%%%%%%%%%%%%%%%%%%%%%%%%%%%%%%%

Let us first focus on the $(\Lambda+1)$th disk with charge $Q$,
which should become infinitely large in the NS5-brane limit.
The potential generated by this disk is given by
\begin{align}
 V_{\Lambda+1}(r,z)=\frac{1}{\pi}\int_{-R}^{R}du\,
 \left(\frac{1}{\sqrt{(z-d+iu)^2+r^2}}-\frac{1}{\sqrt{(z+d+iu)^2+r^2}}\right) f_{\Lambda+1}(u),
\label{pot for Lambda+1}
\end{align}
where $f_{\Lambda+1}(u)$ is the charge density of the highest disk, which is
determined by \eqref{PWMM saddle pt eq}. 
The radius $R$ is related to the charge $Q$ through 
\eqref{PWMM saddle pt eq} and 
\begin{align}
Q=\frac{1}{\pi}\int_{-R}^R du\, f_{\Lambda+1}(u). \label{Q fLambda+1}
\end{align}
%%The total potential is given by \eqref{PWMMpotential}.
%In the NS5 limit, the radius $R$ becomes large.
%As in the NS5 limit for the trivial vacuum of LST,
%a portion of \eqref{pot for Lambda+1} should contribute to make
%the background potential for LST \eqref{simplest NS5potential}.
What we expect in the case of general vacua
is that the potential \eqref{pot for Lambda+1} can be divided into 
two parts: 
one of them contributes to the background part in \eqref{NS5potential}
in the double scaling limit, 
and the other contributes to the rest.
This separation should appear as a result of decomposition of the charge density 
$f_{\Lambda+1}(u)$ as 
% It should be possible to divide the potential \eqref{pot for Lambda+1} into 
% two parts in such a way that, in the double scaling limit, 
% one of them contributes to the background part in \eqref{NS5potential}
% and the other contributes to the rest part.
% This separation should appear as a result of decomposition of the charge density 
% $f_{\Lambda+1}(u)$ as 
%into the part relevant 
%to the background potential \eqref{simplest NS5potential} and the rest,
\begin{align}
 f_{\Lambda+1}(u)=f^{(0)}(u)+g(u),
 \label{f Lambda+1}
\end{align}
where $f^{(0)}(u)$ is the charge density in the case of $\Lambda=0$,
% such that the single conducting disk with radius $R$ is placed at $z=d$.
which satisfies the same equation as \eqref{f0 eq}.
% To be precise, 
We set the support of $f^{(0)}(u)$ to be the same as 
that of $f_{\Lambda+1}(u)$, i.e.~$[-R,R]$,
so that the charge 
$\frac{1}{\pi}\int_{-R}^R du\, f^{(0)}(u)$
is different from $Q$ in \eqref{Q fLambda+1}, instead.
We assume the charge associated with $f^{(0)}(u)$ is of the same order as $Q$ 
when $Q$ is large.
% by adjusting the total charge of the disk, which is different from $Q$ in \eqref{Q fLambda+1}, 
% but of the same order as $Q$ when $Q$ is large.
% and has charge $Q$. 
Meanwhile,
$g(u)$ corresponds to the fluctuation created by the presence 
of the other disks and has support on $[-R,R]$.
Note that, since $g(u)$ is vanishing as $Q_s \to 0$ for any $s$, 
the ratio $g(u)/f^{(0)}(u)$ should be on the order of $\frac{Q_s}{Q}$ 
for large $Q$. % in the limit where $Q$ is large. 
% and $Q_s$ are fixed.
% Note also that, 
% rigorously speaking, the radius of the single disk in defining 
% $f^{(0)}(u)$ is now different from $R$, which is the radius 
% with the presence of the other disks determined from 
% \eqref{PWMM saddle pt eq} and \eqref{Q fLambda+1}.
% So we denote the radius of the single disk by $R'$ in defining 
% $f^{(0)}(u)$. In other words, $f^{(0)}(u)$ is defined to satisfy 
% \begin{align}
% Q=\frac{1}{\pi}\int_{-R'}^{R'}du\, f^{(0)}(u), \label{Q f0}
% \end{align}
% with $f^{(0)}(u)=0$ for $R'<|u|$.
% $R'$ should be of the same order as $R$ and 
% the difference $R'-R$ should become negligible in the limit
%  \eqref{DSLgrav}.
With the decomposition (\ref{f Lambda+1}), 
the argument in the previous subsection ensures that 
the potential generated by $f^{(0)}(u)$ plus 
the background part \eqref{simplest pp-wave potential} produces 
the background potential for LST
in \eqref{simplest NS5potential}, in the NS5-brane limit \eqref{DSLgrav}. 
%(the first term in \eqref{NS5potential}).

% In the following, we assume that $R'<R$.
% This assumption is naturally expected to hold
% %that the charge density $f^{(0)}(u)$ 
% %extends wider than $f_{\Lambda+1}(u)$
% since, in the case of the single disk, 
% there are no other charged disks yielding repulsive forces.
% However, this is just a technical assumption to write down some equations explicitly
% below and one can check that the opposite case with 
% $R< R'$ leads to the same conclusion as far as the difference 
% $R-R'$ becomes negligible in the limit \eqref{DSLgrav}\footnote{This behavior is understood as follows. 
% As $R'$ is the disk radius for the simplest vacuum of PWMM, 
% it satisfies $V_0R'^4/8=Q$ at large $R'$.
% Then the effect from $Q_s$ is supposed to enter in $R$ as
% $V_0R^4/8=Q(1+\mathcal{O}(\frac{Q_s}{Q}))$
% since $R$ becomes equivalent to $R'$ when $Q_s=0$.
% Now the difference between them gets
% $R^4-R'^4 = \mathcal{O}(\frac{1}{V_0})$, which approaches 0 in the limit \eqref{DSLgrav}.}.

We then show that $g(u)$ corresponds to the infinitely many mirror images
in the potential \eqref{NS5potential} for LST.
Note first that $g(u)$ satisfies
% \begin{align}
% 0=\frac{1}{\pi}\int_{-R}^{R}du\, g(u) \label{g}
% \end{align}
% and  
\begin{align}
 g(u)-K_{\Lambda+1}(2d)g(u)-\sum_{t=1}^{\Lambda}(K_t(d+d_t)-K_t(d-d_t))f_t(u)
 =C_{\Lambda+1}-C^{(0)}
 \label{gIntEq}
\end{align}
for $|u|<R$,
% The first equation is obtained by integrating 
% \eqref{f Lambda+1} over $u$ from $-R$ to $R$ and 
% using \eqref{Q fLambda+1} and \eqref{Q f0}.
which follows from \eqref{PWMM saddle pt eq} and \eqref{f0 eq}.
%%%%%%%%%%%%%%%%%%%%%%%%%%%%%%%%%%%%%%%%%%%%%%%%%%%%%%%%%%%%%
%\subsubsection{Potential}
%%%%%%%%%%%%%%%%%%%%%%%%%%%%%%%%%%%%%%%%%%%%%%%%%%%%%%%%%%%%%
Using \eqref{f Lambda+1} and \eqref{gIntEq},
we find that each term in the right-hand side of \eqref{pot for Lambda+1}
can be rewritten as
%Plugging \eqref{gIntEq} into \eqref{pot for Lambda+1} recursively,
%we find that the following relation holds in the limit \eqref{DSLgrav}:
\begin{align}
 &\int_{-R}^{R} du 
 \frac{f_{\Lambda+1}(u)}{\sqrt{(z\mp d+iu)^2+r^2}} 
 \nonumber \\
 &=\int_{-R}^{R} du 
 \frac{f^{(0)}(u)}{\sqrt{(z\mp d+iu)^2+r^2}} 
 % \nonumber \\
 % &\qquad
 \pm \sum_{n=1}^\infty \sum_{s=1}^\Lambda
 \int_{-R_s}^{R_s} du \Bigg(
 \frac{f_s(u)}{\sqrt{(z-d_s\mp 2nd+iu)^2+r^2}}
 \nonumber \\
 &\hspace{90mm}
 -\frac{f_s(u)}{\sqrt{(z+d_s\mp 2nd+iu)^2+r^2}}
 \Bigg) .
 \label{PWMMpotential2parts}
\end{align}
The derivation is shown in appendix \ref{apx:potential_recursion}.
Then, from \eqref{pot for Lambda+1} and \eqref{PWMMpotential2parts}, we obtain
\begin{align}
 V_{\Lambda+1}(r,z)
 &=V^{(0)}(r,z)
 +\sum_{\substack{n=-\infty\\ \neq 0}}^{\infty}\sum_{s=1}^{\Lambda}V_{s,n}(r,z),
% +\lim_{\nu\to \infty}G((2\nu+1)d,r,z),
 \label{PWMMpotential2}
\end{align}
where
\begin{align}
 &V^{(0)}(r,z)=
 \frac{1}{\pi}\int_{-R}^{R} du 
 \left(\frac{1}{\sqrt{(z-d+iu)^2+r^2}}-\frac{1}{\sqrt{(z+d+iu)^2+r^2}}\right) f^{(0)}(u),
 \nonumber \\
 &V_{s,n}(r,z)=
 \frac{1}{\pi}\int_{-R_s}^{R_s} du 
 \left(\frac{1}{\sqrt{(z-d_s-2nd+iu)^2+r^2}}-\frac{1}{\sqrt{(z+d_s-2nd+iu)^2+r^2}}\right) f_s(u).
 \nonumber
\end{align}
Combined with the contributions from the other disks, $V_{s}(r,z)(=V_{s,0}(r,z))\; (s=1,\cdots, \Lambda)$, 
and the background potential for PWMM, in the NS5-brane limit, 
the total electrostatic potential becomes
\begin{align}
V(r,z)=V_0\left(r^2z-\frac{2}{3}z^3\right)+V^{(0)}(r,z)+\sum_{n=-\infty}^{\infty}\sum_{s=1}^{\Lambda}V_{s,n}(r,z).
\label{PWMMpotential3}
\end{align}
In the limit \eqref{DSLgrav}, 
the first two terms in the right-hand side becomes the background potential
for LST \eqref{simplest NS5potential}
with $g_0$ proportional to the fixed quantity in \eqref{DSLgrav}. 
Thus, \eqref{PWMMpotential3} indeed coincides with
the electrostatic potential for a general vacuum of LST \eqref{NS5potential}.
Since the potential of the electrostatic system for PWMM 
becomes completely equivalent to that of LST as a result of
the limit \eqref{DSLgrav}, 
this double scaling limit is indeed the NS5-brane limit, as desired.

%%%%%%%%%%%%%%%%%%%%%%%%%%%%%%%%%%%%%%%%%%%%%%%%%%%%%%%%%%%%%
%\subsubsection{Electrostatic action}
%%%%%%%%%%%%%%%%%%%%%%%%%%%%%%%%%%%%%%%%%%%%%%%%%%%%%%%%%%%%%
We can also understand the NS5-brane limit 
\eqref{DSLgrav} from the action of the 
electrostatic systems. 
% Let us consider the following 
The action of the electrostatic system 
for PWMM\footnote{
Strictly speaking, this is not an action 
but the minus of the potential energy with some Lagrange multipliers.
However, for static configuration, they are equivalent to each other
up to the volume factor of the time direction.
} is as follows,
\begin{align}
 S_{es}&=
 \int_{-\infty}^{\infty}dz \int_0^\infty 2\pi rdr
 \Bigg(
 \frac{1}{8\pi}\nabla V(r,z)\cdot \nabla V(r,z)
 \nonumber \\
 &\qquad\qquad\qquad\qquad\qquad
 -\sum_{s=1}^{\Lambda+1}
 \sigma_s(r)(\delta(z-d_s)-\delta(z+d_s))
 V(r,z)
% \sigma(r)(\delta(z-\kappa)-\delta(z+\kappa)) 
 \Bigg)
 \nonumber \\
 &\qquad\qquad
 +2\sum_{s=1}^{\Lambda+1}C_s\left(
 \int_0^\infty 2\pi rdr\, 
 \sigma_s(r)
 -Q_s
 \right).
\label{ses}
\end{align}
Here, the charge densities are given by
$\sigma_s(r)(\delta(z-d_s)-\delta(z+d_s))$, where
the second term comes from the mirror images.
The coefficients $C_s$ in \eqref{ses} play the role of the Lagrange multipliers
for the condition that the total charge of the $s$th disk is $Q_s$.
The equation of motion for $\sigma_s$ gives the condition that 
the potential on the $s$th disk is constant and is equal to $C_s$.
% The derivation of the action of the electrostatic system for LST 
% from that for PWMM in the NS5-brane limit
% is explained in Appendix \ref{apx:elextrostatic_action}.
In appendix \ref{apx:elextrostatic_action}, 
we show that the action \eqref{ses} reduces to the action 
for LST in the NS5-brane limit \eqref{DSLgrav}.
This gives another ground of \eqref{DSLgrav}.

%%%%%%%%%%%%%%%%%%%%%%%%%%%%%%%%%%%%%%%%%%%%%%%%%%%%%%%%%%%%%
\section{NS5-brane limit in PWMM}
\label{NS5-brane limit in PWMM}
%Emergent bubbling geometry as scalar eigenvalues
%%%%%%%%%%%%%%%%%%%%%%%%%%%%%%%%%%%%%%%%%%%%%%%%%%%%%%%%%%%%%

In this section, we consider the gauge theory side.
%and show that the gravity solutions emerge as the 
%eigenvalue distributions of the adjoint scalar fields in PWMM.
%It is not determined only by the bosonic symmetry,
%but needs the Killing spinor equation.
%We concentrate on PWMM \cite{BMN},
%but the same result for all $SU(2|4)$ symmetric gauge theories 
%can be obtained from the result of PWMM
%because they are described by PWMM 
%as special vacua \cite{Maldacena:2002rb,Ling:2006up,Ishiki:2006yr}.
We first review 
the emergent bubbling geometry from PWMM 
\cite{Asano:2014vba,Asano:2014eca}.
Then, we show that the NS5-brane limit, which we found from the 
gravity side in the last section,
also exists at least in the planar part of
a quarter-BPS sector of PWMM.
In showing this, 
we express \eqref{DSLgrav} in terms of the parameters in PWMM.

\subsection{The plane wave matrix model}

The action of PWMM in the Euclidean signature is given by
\begin{align}
 S&=\frac{1}{g^2}\int d\tau \, \Tr\Bigl(
 \frac{1}{4}F_{MN}F^{MN}
 +\frac{m^2}{2}X_mX^m
 +\frac{i}{2}\Psi \Gamma^M D_M \Psi
 \Bigr).
 \label{action of PWMM}
\end{align}
We use the same notation as in \cite{Asano:2014vba,Asano:2014eca}.
The indices $M, N$ run from $1$ to $10$. 
They are decomposed as $M= (1,a,m)$, where $a$ runs from $2$ to $4$
and $m$ runs from $5$ to $10$.
$F_{MN}$ and $D_M \Psi $ are given by 
\begin{align}
&F_{1M}=D_1X_M = \partial_1 X_M -i[X_1, X_M]  \;\; (M \neq 1),
\nonumber\\
& F_{ab}=m\epsilon_{abc}X_c-i [X_a, X_b], \;\; 
F_{am}=-i[X_a,X_m], \;\;
F_{mn}=-i[X_m,X_n],
\nonumber\\
& D_1 \Psi = \partial_1 \Psi -i[X_1, \Psi], \;\;
D_a \Psi = \frac{m}{8}\epsilon_{abc}\Gamma^{bc}\Psi
-i[X_a, \Psi], \;\; 
D_m\Psi = -i[X_m , \Psi].
\end{align}
$\partial_1$ is the derivative with respect to $\tau$
and $X_1$ is the 1-dimensional gauge field.
Throughout this paper, 
the mass parameter $m$ is set to 2,
without loss of generality.

PWMM has the $SU(2|4)$ symmetry.
The vacua of PWMM are given by fuzzy spheres, which preserve
the entire $SU(2|4)$ symmetry
and are labeled by the $SU(2)$ representations.
The vacuum configuration is given by $X_m =0$ and 
\begin{align}
 X_a=-2L_a=-2\bigoplus_{s=1}^{\Lambda +1}\bm{1}_{N_2^{(s)}}\otimes L_a^{[D_s]}
 \qquad (a=2,3,4),
 \label{fuzzy sphere}
\end{align}
where $\Lambda$, $N_2^{(s)}$ and $D_s$ are integers and 
$L_a^{[D]}$ stands for the $SU(2)$ generators in the $D$ dimensional 
irreducible representation. They satisfy $[L_a^{[D]}, L_b^{[D]}] 
=i \epsilon_{abc}L_c^{[D]}$.
The matrix size of PWMM is given by 
$\sum_{s=1}^{\Lambda+1}N_2^{(s)}D_s$.
The right-hand side of (\ref{fuzzy sphere})
is just the irreducible decomposition of a general 
reducible representation. 

The parameters labeling the vacua are the 
multiplicities and dimensions of irreducible representations,
$\{ N_2^{(s)},D_s\}_{s=1,\cdots,\Lambda+1}$.
Note that we have already used the same names for variables 
on the gravity side. 
We used $N_2^{(s)}$ for the D2-brane charges 
and $D_s$ for the position of the disks, which is related 
to the NS5-brane charges.
This usage is justified since
these variables can be indeed identified with 
each other through the gauge/gravity correspondence.
The identification for $N_2^{(s)}$ is 
easily understood because it corresponds 
to the number of the fuzzy spheres in the vacuum 
(\ref{fuzzy sphere}), so that it should be equal to the 
D2-brane charges. 
% The identification for $D_s$ is rather non-trivial. 
% By comparing the mass spectra of PWMM and of a
% spherical fivebrane in the supergravity approximation,  
% it was shown that a general vacuum of PWMM corresponds to 
% a state with fivebranes, where the dimensions of the 
% irreducible representations $D_s$ are related to 
% the fivebrane charges by $D_s=\sum_{t=1}^{s}N_5^{(t)}$, 
% which is exactly the same relation we had on the gravity side
% \cite{Maldacena:2002rb}.
%%%
The identification for $D_s$ can be understood as follows.
The UV region of the gravity solution \eqref{LM solution} becomes the D0-brane 
solution, where the D0-brane charge is proportional to 
the dipole moment, $2\sum_s D_s Q_s$, 
generated by the conducting disks and mirror images. 
From the above identification 
of the D2-brane charges and the fact that the D0-brane charge 
is equal to the matrix size of PWMM, one finds that $D_s$ correspond to 
the dimensions of the irreducible representations\footnote{
This identification can also be understood from the M-theoretic 
point of view. By comparing the mass spectra of PWMM and of a spherical 
M5-brane in the supergravity approximation, it was argued that a 
general vacuum of PWMM corresponds to a state with fivebranes,
where the number of fivebranes is given by the dimension of 
the largest irreducible representation \cite{Maldacena:2002rb}. 
This relation is consistent with 
the identification for $D_s$, which states that 
$D_{\Lambda+1}$ is indeed given by the sum over all fivebrane charges.}.
%%%
For each theory around a vacuum labeled by
$\{ N_2^{(s)},D_s\}_{s=1,\cdots,\Lambda+1}$, 
the dual geometry on the gravity side is given by the 
Lin-Maldacena geometry with the same parameters. 

%When one writes the geometry in terms of an electrostatic system,
%$N_2^{(s)}$ and $D_s$ correspond to 
%the electric charge on the $s$th disk \eqref{D2 charge}
%and the $z$-coordinate of the disk \eqref{NS5 charge},
%respectively. 

\subsection{Emergent geometries}
If the gauge/gravity correspondence holds true, the dual geometry 
should be somehow realized in the strong coupling limit of PWMM.
In order to see the emergence of the dual geometry, we apply the localization method, 
which makes exact calculations possible for some supersymmetric sectors
\cite{Pestun:2007rz}.

We first choose an appropriate BPS sector.
Note that the emergence of $S^2$ and $S^5$ in 
the geometry (\ref{LM solution}) 
is rather trivial because any vacua of PWMM 
preserve $SO(3)\times SO(6)$ symmetry.
So the only non-trivial part is the emergence of the space 
parametrized by $r$ and $z$.
It was found that the vacuum expectation values (VEVs) of operators made of the complex scalar field
\begin{align}
 \phi (\tau) = -X_4(\tau)+ X_9(\tau)\sinh \tau + i X_{10}(\tau)\cosh \tau,
 \label{phi PWMM}
\end{align}
describe the $r$- and $z$-directions\footnote{
% It looks contradictory that $\phi$ is not $SO(3)\times SO(6)$ invariant
% while $r$ and $z$ are.
% This is understood as follows:
By taking VEVs, % of $\phi$, 
one obtains the $SO(3)\times SO(6)$ invariant part of $\phi$,
which should contain the information of the $r$- and $z$-directions.
}\cite{Asano:2014vba,Asano:2014eca}. 
This can be seen as follows. If we make a Wick-rotation to go back 
to the Lorentzian signature, the real and imaginary parts of 
$\phi$ at each fixed time are given by an $SO(3)$ scalar 
and an $SO(6)$ scalar, respectively, up to an $SO(2)$ ($\subset SO(6)$) rotation. 
So this field corresponds to a point on $S^2 \times S^5$ 
in the dual geometry and thus corresponds to the perpendicular directions,
that is, the space of $r$ and $z$.
On the other hand, when $X_{10}$ direction is further Wick-rotated while 
the time is kept to be Euclidean, the field  (\ref{phi PWMM}) is shown to
preserve a quarter of the whole supersymmetry 
and the vacuum expectation value % (VEV) 
of any operators made only of this field 
can be computed by the localization method.

The localization computation for 
the partition function and the VEV of such operators 
was done in \cite{Asano:2012zt}.
Let us review this computation.
One first needs to impose a boundary condition 
to define the theory
as the time direction of PWMM is not compact.
A natural condition for the theory around a fixed vacuum is 
that all fields approach the vacuum configuration 
as $\tau\to\pm\infty$. 
After imposing this condition so that the theory is defined appropriately,
one adds a $Q$-exact term to the action, where $Q$ stands for 
a supersymmetry transformation which keeps $\phi$ invariant.
Then, from the standard argument of the localization computation \cite{Pestun:2007rz},
one-loop calculations around the saddle point of the $Q$-exact term 
give an exact result.
At the saddle point, 
all fields except $X_{10}$ take the vacuum configuration 
while there are non-trivial moduli for $X_{10}$. 
The saddle point configuration is written as
\begin{align}
 X_a(\tau)=-2L_a
 \qquad (a=2,3,4),
 \qquad
 X_{10}(\tau)=\frac{M}{\cosh\tau},
\label{saddle config}
\end{align}
where all the other fields are vanishing at the saddle point. 
$M$ is a constant Hermitian matrix commuting with $L_a$'s.
When $L_a$ are decomposed as in the right-hand side of 
\eqref{fuzzy sphere},
$M$ can be decomposed as
$M=\bigoplus_{s=1}^{\Lambda +1}M_s\otimes\mathbf{1}_{D_{s}}$,
where $M_s$ is an $N_2^{(s)}\times N_2^{(s)}$ Hermitian matrix.
After calculating the one-loop determinant around the saddle point
\eqref{saddle config},
one obtains the following form for the partition function \cite{Asano:2012zt}:
\begin{align}
 &Z_{{\cal R}}=\int \prod_{s=1}^{\Lambda+1}
 \prod_{i=1}^{N_2^{(s)}}dm_{si}\, Z_{\rm 1-loop}({\cal R},\{m_{si}\})\,
 e^{-\frac{2}{g^2}\sum_{s}\sum_{i}D_sm_{si}^2},
 \label{matrix model}
\end{align}
where ${\cal R}$ stands for the representation of (\ref{fuzzy sphere})
labeled by $\{ N_2^{(s)},D_s\}_{s=1,\cdots,\Lambda+1}$ and
$m_{si}$'s are eigenvalues of $M_s$.
$Z_{\rm 1-loop}({\cal R},\{m_{si}\})$ is the one-loop determinant 
around the saddle point 
%with the representation ${\cal R}$
given by
\begin{align}
 Z_{\rm 1-loop}=
 \prod_{s,t=1}^{\Lambda+1}
 %\prod_{J=|D_s-D_t|/2}^{(D_s+D_t)/2-1}
 \prod_{J}
 \prod_{i=1}^{N_2^{(s)}}\prod_{j=1}^{N_2^{(t)}}
 \hspace{-5.5mm} {\phantom{\prod}}^{\prime}
 \left[
 \frac{\{(2J+2)^2+(m_{si}-m_{tj})^2\} \{(2J)^2+(m_{si}-m_{tj})^2\}}
 {\{(2J+1)^2+(m_{si}-m_{tj})^2\}^2}
 \right]^{\frac{1}{2}}.
 \label{1loopdet}
\end{align}
Here $J$ runs from $|D_s-D_t |/2$ to $(D_s+D_t)/2-1$.
We denote by $\prod'$ the product in which 
the second factor in the numerator with 
$s=t$, $J=0$ and $i=j$ is not included.
%Although this is not the entire partition function of PWMM but a partial one,
%it is well-defined in the large-$N$ limit 
%due to the suppress of instanton effect.
% For the VEV of operators made only of $\phi$, 
% say $f(\phi)$,
% the following relation holds,
Then,
any operator in the BPS sector, which is made only of $\phi$,
can be computed by the partition function \eqref{matrix model};
namely, 
the VEV of a function of $\phi$, say $f(\phi)$,
is computed as
\begin{align}
 \langle f(\phi) \rangle = \langle f(-2L_4+ iM) \rangle_M,
\end{align}
where $\langle \cdots \rangle_M$ stands for the expectation value with 
respect to the partition function (\ref{matrix model}).

The partition function (\ref{matrix model}) describes 
the quarter-BPS sector of the theory around the vacuum labeled by 
the representation ${\cal R}$.
Insertions of operators of $\phi$ 
correspond to the quarter-BPS fluctuations around the vacuum.
On the supergravity side, they will correspond to the fluctuations 
of fields which preserve the same supersymmetry.
On the other hand, if there is no insertion of the operators, 
the partition function (\ref{matrix model})
is dominated by the vacuum configuration, 
which should contain the information 
of the background geometry of the dual string theory.
If we can solve the model (\ref{matrix model}) completely, 
we can check whether these correspondences are true or not. 
Though solving this model for arbitrary parameters 
is very difficult, 
the problem is simplified in the large-$N$ limit, where 
the eigenvalue integral can be evaluated
by the saddle point approximation.
This is also the regime where the supergravity 
approximation becomes valid.
% on the gravity side.

An appropriate parameter 
region to see the correspondence
is given as follows \cite{Asano:2014vba,Asano:2014eca}.
In order to suppress the bulk string coupling correction,
we take the 't~Hooft limit
\begin{align}
 N_2^{(s)}\rightarrow \infty, \quad \lambda^{(s)}=g^2N_2^{(s)}=\text{fixed}.
 \label{tHooft limit}
\end{align}
Moreover, to suppress the $\alpha'$ corrections, we consider the region where
\begin{align}
 N_5^{(s)}=D_s-D_{s-1} \gg 1, 
 \quad 
 \lambda^{(s)}\gg D_s,
 \label{limit}
\end{align}
for arbitrary $s$. 
In this parameter region, 
the effective action of \eqref{matrix model} is given by
\begin{align}
 S_{eff}
 &=\sum_{s=1}^{\Lambda+1} \frac{2D_s}{g^2} \int dx \ x^2 \rho^{(s)}(x)
 +\sum_{s=1}^{\Lambda+1}\frac{\pi}{2}  \int dx\, \rho^{(s)}(x)^2
 \nonumber \\
 &\quad
 -\sum_{s,t=1}^{\Lambda+1}\frac{1}{2} \int dx dy \left[
 \frac{D_s+D_t}{(D_s+D_t)^2+(x-y)^2}
 -\frac{|D_s-D_t|}{(D_s-D_t)^2+(x-y)^2}\right] \rho^{(s)}(x)\rho^{(t)}(y) 
 \nonumber \\
 &\quad
 -\sum_{s=1}^{\Lambda+1}\mu_s\left(\int dx\, \rho^{(s)}(x)-N_2^{(s)}\right),
 \label{effective action}
\end{align}
where $\rho^{(s)}(x)$ is the eigenvalue distribution,
$\rho^{(s)}(x)=\sum_{i=1}^{N_2^{(s)}}\delta(x-m_{si})$, 
and $\mu_s$ is the Lagrange multiplier imposing the normalization for
$\rho^{(s)}(x)$:
\begin{align}
N_2^{(s)}=\int_{-x^{(s)}_m}^{x_m^{(s)}} dx\, \rho^{(s)}(x).
\label{rho normalization}
\end{align}
$x_m^{(s)}$ is the extent of the support of $\rho^{(s)}(x)$.
In the 't~Hooft limit, the saddle point approximation for the variables
$\rho^{(s)}$ becomes exact and the theory becomes basically a classical theory with the action (\ref{effective action}).

Let us now show that this classical theory is equivalent to the 
axially symmetric electrostatic system considered on the gravity side.
Below, we verify this equivalence\footnote{
This equivalence was 
originally argued by using the equations 
of motion in \cite{Asano:2014vba,Asano:2014eca}.
} by comparing the electrostatic action (\ref{ses}) and 
the quarter-BPS effective action (\ref{effective action}).
%% To this end, we make a change of the parameters and the variables
%% in (\ref{effective action}) as
%% \begin{align}
%%  &\frac{1}{g^2}=\frac{\pi^2}{2}V_0, \label{g V0}\\
%%  &\rho^{(s)}(x)=\frac{4}{\pi^2}f_s\left(\frac{\pi}{2}x\right), \label{rho f}\\
%%  &x_m^{(s)}=\frac{2}{\pi}R_s , \label{x R}\\
%%  &\mu_s=\frac{4}{\pi}(C_s+\frac{2}{3}V_0d_s^3),
%%  \label{iden}
%% \end{align}
%% where in the third equation, $x_m^{(s)}$ stands for 
%% the upper edge of the support of
%% the eigenvalue distribution $\rho^{(s)}(x)$.
%% Then we rewrite the action (\ref{effective action}) using 
%% the quantities on the right-hand sides in (\ref{g V0} -- \ref{iden}).
%% By substituting them into \eqref{effective action},
%% we obtain 
%% \begin{align}
%% S_{eff}
%%  &=\frac{32}{\pi^4}\Bigg[
%%  \sum_{s=1}^{\Lambda+1} 2V_0d_s \int_{-R_s}^{R_s} du \ u^2 f_{s}(u)
%%  +\sum_{s=1}^{\Lambda+1}\frac{1}{2}  \int du\, f_{s}(u)^2
%%  \nonumber \\
%%  &\quad
%%  -\sum_{s,t=1}^{\Lambda+1}\frac{1}{2\pi} \int du du' \left[
%%  \frac{d_s+d_t}{(d_s+d_t)^2+(u-u')^2}
%%  -\frac{|d_s-d_t|}{(d_s-d_t)^2+(u-u')^2}\right] f_{s}(u)f_{t}(u') 
%%  \nonumber \\
%%  &\quad
%%  -\sum_{s=1}^{\Lambda+1}(C_s+\frac{2}{3}V_0d_s^3)\left(\int du\, f_{s}(u)-\pi Q_{s}\right)
%%  \Bigg].
%% \label{seff1}
%% \end{align}
%% %We claim that this action is essentially equivalent to the standard action 
%% %for the axially symmetric electrostatic system \eqref{ses}.

Back on the gravity side,
we first rewrite the electrostatic action (\ref{ses}) in terms of the charge densities $f_s(u)$. 
By using \eqref{PWMMpotential} and \eqref{fandsigma}, one can show that 
the potential $V_s$'s satisfy the following relations\footnote{
It is convenient to use the following expression for $V(r,z)$,
\begin{align}
 V_s(r,z)
 &=
 4\pi
 \int dz' r'dr' \, G(r,z;r',z')\sigma_s(r')(\delta(z'+d_s)-\delta(z'-d_s)).
\nonumber
 % \nonumber \\
 % &=
 % \frac{1}{\pi}\int_{-R_s}^{R_s}du\left( 
 % -\frac{1}{\sqrt{r^2+(z+d_s+iu)^2}}+\frac{1}{\sqrt{r^2+(z-d_s+iu)^2}}
 % \right) f_s(u),
\end{align}
Here,
\begin{align}
 G(r,z;r',z')=-\frac{1}{2}\int_{0}^{\infty} du\, e^{-u|z-z'|}J_0(ru)J_0(r'u) \nonumber
\end{align}
is the Green function of the Laplacian in the cylindrical coordinate, where
$J_0$ is the Bessel function of the 0th order.},
\begin{align}
 &\int_{-\infty}^\infty dz\int_0^{\infty} rdr\, 
  V_s(r,z)\bigtriangleup  V_t(r,z)
 \nonumber \\
 % &=
 % \int_{-\infty}^\infty dz\int_0^{\infty} rdr
 % (-4\pi)\sigma_t(r) (\delta(z-d_t)-\delta(z+d_t)) V_s(r,z)
 % \nonumber \\
 % &=-16\pi^2
 % \int_{-\infty}^\infty dz\int_0^{\infty} rdr
 % \int_{-\infty}^\infty dz'\int_0^{\infty} r'dr'
 % \nonumber \\
 % &\qquad
 % \sigma_t(r)(\delta(z-d_t)-\delta(z+d_t))
 % G(r,z;r',z')
 % \sigma_s(r')(\delta(z'+d_s)-\delta(z'-d_s))
 % \nonumber \\
 &\quad
 =
 \frac{4}{\pi^2} \bigg(
 \int_{-\infty}^{\infty}dudu'
 \frac{|d_s+d_t|}{(d_s+d_t)^2+(u-u')^2}f_s(u)f_t(u')
 \nonumber \\
 &\qquad\qquad
 -\int_{-\infty}^{\infty}dudu'
 \frac{|d_s-d_t|}{(d_s-d_t)^2+(u-u')^2}f_s(u)f_t(u')
 \bigg) \;\;\;  (t\neq s),\\
 &\int_{-\infty}^\infty dz\int_0^{\infty} rdr\, 
  V_s(r,z)\bigtriangleup  V_s(r,z)
 \nonumber \\
 &\quad
 =
 \frac{4}{\pi^2} \bigg(
 \int_{-\infty}^{\infty}dudu'
 \frac{2d_s}{(2d_s)^2+(u-u')^2}f_s(u)f_s(u')
 % \nonumber \\
 % &\qquad\quad
 -\pi \int_{-\infty}^{\infty}du
 f_s(u)^2
 \bigg),\\
 &\int_{-\infty}^{\infty}dz
 \int_0^\infty rdr\, (r^2z-\frac{2}{3}z^3)\bigtriangleup V_s(r,z)
 =
 \frac{4}{\pi}\int_{-1}^1 du
 \left( -2d_s u^2+\frac{2d_s^3}{3}\right) f_s(u).
\end{align}
By using these relations, we can rewrite \eqref{ses} as
\begin{align}
S_{es}&=-\frac{2}{\pi}\Bigg[
 \sum_{s=1}^{\Lambda+1} 2V_0d_s \int_{-R_s}^{R_s} du \ u^2 f_{s}(u)
  +\sum_{s=1}^{\Lambda+1}\frac{1}{2}  \int du\, f_{s}(u)^2
  \nonumber \\
  &\qquad
  -\sum_{s,t=1}^{\Lambda+1}\frac{1}{2\pi} \int du du' \left[
  \frac{d_s+d_t}{(d_s+d_t)^2+(u-u')^2}
  -\frac{|d_s-d_t|}{(d_s-d_t)^2+(u-u')^2}\right] f_{s}(u)f_{t}(u') 
  \nonumber \\
  &\qquad
  -\sum_{s=1}^{\Lambda+1}(C_s+\frac{2}{3}V_0d_s^3)\left(\int du\, f_{s}(u)-\pi Q_{s}\right)
  \Bigg], \label{ses f}
% \nonumber \\
%% &\quad
%% +\frac{4 V_0}{3}\sum_{s=1}^{\Lambda+1}d_s^3Q_{s}
%%  +\frac{1}{4}
%%  \int_{-\infty}^{\infty} dz \int_0^\infty rdr
%%  \bigg(
%%  \nabla V_{bg}(r,z)\cdot \nabla V_{bg}(r,z)
%%  \nonumber \\
%%  &\qquad\qquad\qquad\qquad\qquad\qquad\qquad\qquad
%%  +\nabla \cdot \{  V(r,z)\nabla (2V_{bg}(r,z)+ V(r,z))\} \bigg).
\end{align}
where we neglected total derivatives and terms that contain no dynamical variables.

Comparing the gauge-theory side \eqref{effective action} and 
the gravity side \eqref{ses f}, we find the equivalence relation
\begin{align}
S_{es}=-\frac{\pi^3}{16}S_{eff}
\label{equivalence of actions}
\end{align}
under the identifications
\begin{align}
 &\frac{1}{g^2}=\frac{\pi^2}{2}V_0, \label{g V0}\\
 &\rho^{(s)}(x)=\frac{4}{\pi^2}f_s\left(\frac{\pi}{2}x\right), \label{rho f}\\
 &x_m^{(s)}=\frac{2}{\pi}R_s , \label{x R}\\
 &\mu_s=\frac{4}{\pi}(C_s+\frac{2}{3}V_0d_s^3).
 \label{iden}
\end{align}
%% \begin{align}
%% S_{eff}
%%  &=\frac{4}{\pi^3} \int_{-\infty}^{\infty}dz  \int_{0}^{\infty}rdr
%%  \Bigg[
%%  -\sum_{s,t=1}^{\Lambda+1} V_s(r,z)\bigtriangleup  V_t(r,z)
%%  \nonumber \\
%%  &\qquad
%%  -2\sum_{s=1}^{\Lambda+1}\left( 
%%  V_0(r^2z-\frac{2}{3}z^3)-C_s\frac{z}{d_s}
%%  \right) \bigtriangleup  V_s(r,z)
%%  \Bigg]
%%  +\frac{32}{\pi^3}\sum_{s=1}^{\Lambda+1}(C_s+\frac{2}{3}V_0d_s^3)Q_{s}.
%%  \label{electrostatic eff action}
%% \end{align}
%% This is related to \eqref{ses} as 
%% %This is the same as the standard electrostatic action (\ref{ses})
%% %up to irrelevant constant terms and total derivatives.
%% %The relation is explicitly given by
%% \begin{align}
%%  S_{es}
%%  &=
%%  -\frac{\pi^3}{16}S_{eff}
%%  +\frac{4 V_0}{3}\sum_{s=1}^{\Lambda+1}d_s^3Q_{s}
%%  +\frac{1}{4}
%%  \int_{-\infty}^{\infty} dz \int_0^\infty rdr
%%  \bigg(
%%  \nabla V_{bg}(r,z)\cdot \nabla V_{bg}(r,z)
%%  \nonumber \\
%%  &\qquad\qquad\qquad\qquad\qquad\qquad\qquad\qquad
%%  +\nabla \cdot \{  V(r,z)\nabla (2V_{bg}(r,z)+ V(r,z))\} \bigg).
%%  \label{es and eff}
%% \end{align}
%% The second and the third terms on the right-hand side 
%% do not contain the dynamical variables and the fourth term is a
%% total derivative. 
Therefore, $S_{es}$ and $S_{eff}$ are equivalent up to irrelevant constant terms and total derivatives,
and thus give the same extremum.
Some consistency checks of this identification can be found in 
\cite{Asano:2014vba,Asano:2014eca}.
It turns out from \eqref{rho f} that the $s$th eigenvalue distribution $\rho^{(s)}(x)$ % of the scalar field 
in PWMM
is equivalent to the charge density on the $s$th disk in the dual gravity.
Note that the charge density fully determines the geometry
once the gravity solution is in the form of \eqref{LM solution}
with the Laplace equation~\eqref{axi-Laplace-eq} and
the positivity and regularity conditions.
In this sense, one can say that the dual geometry is formed by 
the eigenvalue distribution of the scalar fields in PWMM, 
through the relation (\ref{rho f}).
Therefore, this provides 
a glimpse into a manifestation 
% of the emergence
of geometry from PWMM.

%%%%%%%%%%%%%%%%%%%%%%%%%%%%%%%%%%%%%%%%%%%%%%%%%%%%%%%%%%%%%
\subsection{The double scaling limit in PWMM}
%%%%%%%%%%%%%%%%%%%%%%%%%%%%%%%%%%%%%%%%%%%%%%%%%%%%%%%%%%%%%

The equivalence \eqref{equivalence of actions} 
clearly shows that the planar part of the quarter-BPS sector of PWMM 
has the same NS5-brane limit as the gravity side discussed in
section \ref{section NS5-brane limit}.
%describes LST on $R\times S^5$ in the NS5 limit. 
%The arguments presented so far show the possibility that PWMM describes LST on $R\times S^5$ 
%in the NS5 limit \eqref{DSLgrav}. 
In this subsection, we rewrite the NS5-brane limit 
\eqref{DSLgrav} using the parameters on the gauge theory side:
$D:=D_{\Lambda+1}$, $N_2:=N_2^{(\Lambda+1)}$ and $g$.
%by rewriting the NS5 limit \eqref{DSLgrav} in terms of the quantities in PWMM.

To do this, we have to rewrite
$d$ ($=d_{\Lambda+1}$), $R$ ($=R_{\Lambda+1}$) 
and $V_0$
in \eqref{DSLgrav} in terms of the quantities in PWMM.
% In the following, we again use the notation $d_{\Lambda+1}=:d$ and 
% $R_{\Lambda+1}=:R$.
It is already shown that $d$ and $V_0$ are given in terms of $D$ and $g^2$, respectively, 
in \eqref{brane charge} and \eqref{g V0},
whereas $R$ is given by the upper edge of the support 
($x_m^{(\Lambda+1)}$ in \eqref{x R}) of the 
eigenvalue density $\rho^{(\Lambda+1)}(x)$.
Below, we will 
partly solve the saddle point equations for the eigenvalue densities and express $x_m^{(\Lambda+1)}$ explicitly in terms of the parameters of PWMM.

%First, let us find what $R$ corresponds to on the gauge theory side.
%Let us rewrite the $(\Lambda+1)$th quantities in this subsection as
%$x_m^{(\Lambda+1)}=x_m$, $\rho^{(\Lambda+1)}(x)=\rho (x)$, 
%$N_2^{\Lambda+1}=N_2$ and $D_{\Lambda+1}=D$ for simplicity.
%% The saddle point equation of \eqref{matrix model} 
%% in the supergravity approximation can be written as
The saddle point equation of $\rho^{(s)}(x)$ 
is obtained from \eqref{effective action} as
\begin{align}
 \mu_{s}=
 \frac{2D_{s}}{g^2}x^2+\pi \rho^{(s)}(x)
 -\frac{i}{2}\sum_{t=1}^{\Lambda +1}
 \Big\{ \big( \omega^{(t)}(x+i(D_s+D_t))-\omega^{(t)}(x-i(D_s+D_t)) \big)
 \nonumber \\
 -\big( \omega^{(t)}(x+i|D_s-D_t|)-\omega^{(t)}(x-i|D_s-D_t|) \big) \Big\},
 % \mu_{s}=\frac{2D_{s}}{g^2}x^2+
 % \sum_{t=1}^{\Lambda+1}
 % \sum_{J=\frac{|D_s-D_t|}{2}}^{\frac{D_s+D_t}{2}-1}
 % \int dy\, \rho^{(t)}(y)
 % \log \frac{[(2J)^2+(x-y)^2][(2J+2)^2+(x-y)^2]}{[(2J+1)^2+(x-y)^2]^2},
\label{saddle pt eq rho}
\end{align}
where $\omega^{(t)}(z)$ are the resolvents for each eigenvalue density, 
$\omega^{(t)}(z):=\int dx\, \rho^{(t)}(x)/(z-x)$.
$\rho^{(t)}(x)$ is obtained from $\omega^{(t)}(z)$ via
$\rho^{(t)}(x)=-\frac{1}{2\pi}(\omega^{(t)}(x+i0)-\omega^{(t)}(x-i0))$.
Now, let us consider the saddle point equations for $\rho^{(\Lambda+1)}(x)$ in the NS5-brane limit \eqref{DSLgrav}. 
In terms of $D_s$ and $x_m^{(\Lambda+1)}$, the limit 
implies $x_m^{(\Lambda+1)}\gg D_s$.
Then, the typical scale of the eigenvalue density $\rho^{(\Lambda+1)}(x)$ is naturally expected 
as $|x|\sim x_m^{(\Lambda+1)} \gg D_s$.
In this regime, the saddle point equations \eqref{saddle pt eq rho} are reduced to a much simpler form:
\begin{align}
 \mu_{s} = \frac{2D_{s}}{g^2}x^2
 +\left(D_{s}\sum_{t\geq s}+\sum_{t< s}D_{t}\right) 
 \big( \omega^{(t)\prime}(x+i0)+\omega^{(t)\prime}(x-i0) \big) .
\end{align}
From the equations for $s=\Lambda$ and $\Lambda+1$, one obtains the relation
\begin{align}
 \bar\mu -\frac{2}{g^2}x^2
 =\omega^{(\Lambda+1)\,\prime}(x+i0)+\omega^{(\Lambda+1)\,\prime}(x-i0), \label{omegaprime eq}
\end{align}
where $\bar\mu=(\mu_{\Lambda+1}-\mu_{\Lambda})/N_5^{(\Lambda+1)}$.
% and $\omega (z)=\omega^{(\Lambda+1)}(z)$.
% \begin{align}
%  \frac{\mu_{\Lambda+1}-\mu_{\Lambda}}{N_5^{(\Lambda+1)}-N_5^{(\Lambda)}}
%  -\frac{2}{g^2}x^2
%  =\omega^{(\Lambda+1)\prime}(x+i0)+\omega^{(\Lambda+1)\prime}(x-i0)
% \end{align}
As shown in appendix \ref{apx:solution}, integrating \eqref{omegaprime eq} over $x$ yields 
the saddle point equation for a well-known quartic matrix model, which can be easily solved.
As a result, the eigenvalue density for $\omega^{(\Lambda+1)}(z)$ is obtained as
\begin{align}
 \rho (x)
 =\frac{x_m}{2\pi} \Bigg( 
 \bar\mu -\frac{2x_m^2}{3g^2}\left( \frac{1}{2}+\frac{x^2}{x_m^2}\right)
 \Bigg) \left( 1-\left( \frac{x}{x_m} \right)^2 \right)^{\frac{1}{2}},
 \label{rho not critical}
\end{align}
where $\rho (x)=\rho^{(\Lambda+1)}(x)$ and $x_m=x_m^{(\Lambda+1)}$.
The normalization of $\rho(x)$ \eqref{rho normalization} imposes a condition among parameters:
\begin{align}
 \bar\mu
 =\frac{x_m^{4}+8g^2N_2}{2g^2x_m^2}.
 \label{bar_mu}
\end{align}

Recall that there is one more condition for $\rho(x)$.
As mentioned in section \ref{dual geometries of PWMM and LST}, 
the finiteness of the corresponding gravity solution requires 
each charge density $f_s(u)$ to have a finite derivative at the edges.
Imposing the same condition on $\rho(x)$ together with \eqref{bar_mu},
%on the gauge theory side that
%the derivative of the eigenvalue density with respect to $x$ 
%is zero at the end of the support,
we obtain
% \begin{align}
%  &x_m^{2}
%  =g^2\bar\mu .
% \end{align}
\begin{align}
 x_m^{4}
 =8g^2N_2.
 \label{xm^4}
\end{align}
This relation can be also derived from purely
field theoretic view point by application of the least action principle. 
The detail of this derivation is explained in appendix \ref{apx:solution}.
%With the relation \eqref{x R}, one finds that the radius of the highest 
%disk $R$ is given as 
%\begin{align}
%R=\frac{\pi}{2}(8g^2N_2)^{1/4}.
%\end{align}
Then the eigenvalue density \eqref{rho not critical} 
%with \eqref{bar_mu} and \eqref{xm^4} 
becomes
\begin{align}
 \rho (x)
 =\frac{x_m^3}{3\pi g^2}
 \left( 1-\left( \frac{x}{x_m} \right)^2 \right)^{\frac{3}{2}}.
 \label{rho critical}
\end{align}
This is exactly the same form as that for $\Lambda=1$ given by 
\eqref{critical f0}.

Finally, by substituting the relations \eqref{brane charge}, \eqref{g V0} and \eqref{xm^4}
to \eqref{DSLgrav},
we find that the NS5-brane limit can be written as
\begin{align}
 N_2\to \infty,\quad 
 \frac{1}{N_2}(g^2N_2)^{\frac{5}{8}}
 \exp \left[
 {2^{\frac{3}{4}}\pi \frac{(g^2N_2)^{\frac{1}{4}}}{D}}
 \right]
 \equiv \tilde g_s\text{ : fixed}.
\label{DSLgauge}
\end{align}
% where $N_2 = N_2^{(\Lambda+1)}$ and $D=D_{\Lambda+1}$.
The fixed parameter $\tilde g_s$ corresponds to $g_0$ in \eqref{simplest NS5potential} on the gravity side,
and hence, it is argued in \cite{Ling:2006up} that
$\tilde g_s$ is considered as an 
effective coupling of the IIA LST.
%Note that, as we have shown the direct correspondence between the eigenvalue densities and the charge densities,
%this NS5 limit exists for any quarter BPS operators made of $\phi(\tau)$ in 
%PWMM.

At finite $N_2$,
there are possible non-planar corrections to 
the quantities \eqref{rho f}--\eqref{iden},
which could correct the fixed parameter in the expression \eqref{DSLgauge}.
% could be corrected as well.
However, we claim that the double scaling limit \eqref{DSLgauge} should be valid
with the non-planar corrections taken into account.
% which we denote by $\tilde{\tilde g}_s=\tilde g_s+\delta\tilde g_s$.
% If the ratio of the correction $\delta\tilde g_s$ to $\tilde g_s$
% approaches $\delta\tilde g_s/\tilde g_s\to 0$ for large $N_2$,
% the fixed parameter remains uncorrected.
Obviously, the correction appears
when the ratio of the correction $\delta\tilde g_s$ to the original $\tilde g_s$
approaches a finite value or infinity in the double scaling limit \eqref{DSLgauge}.
In the former case,
each $g$-dependent coefficient in the non-planar corrections
appears such that the coefficient combines with $1/N_2$
to form $\tilde g_s$ in the limit.
Thus, the corrected fixed parameter $\tilde g_s+\delta\tilde g_s$ becomes
a polynomial in $\tilde g_s$,
and one can still fix $\tilde g_s$ for the double scaling limit.
The latter case should be impossible if the gauge/gravity duality holds
because the perturbative expansion in terms of quantum correction
breaks down otherwise.
Therefore, the double scaling limit should remain unchanged
by the non-planar corrections.

%%%%%%%%%%%%%%%%%%%%%%%%%%%%%%%%%%%%%%%%%%%%%%%%%%%%%
\section{Numerical results}\label{sec:numerical_results}
%%%%%%%%%%%%%%%%%%%%%%%%%%%%%%%%%%%%%%%%%%%%%%%%%%%%%
In this section, we show our numerical results for the quarter-BPS sector in PWMM
and discuss the existence of the double scaling limit.
We employ the hybrid Monte Carlo method to perform 
the integration in the partition function \eqref{matrix model},
which is the effective theory of the quarter-BPS sector.
Since the numerical results provide information of $N_2$-dependence,
one can discuss not only the existence of the double scaling limit
but also quantum loop correction, which
is interpreted as $\tilde g_s$ correction in the LST
as we will see below.

%%%%%%%%%%%%%%%%%%%%%%%%%%%%%%%%%%%%%%%%%%%%%%%%%%%%%
\subsection{Expansion with respect to the LST coupling}

The double scaling limit \eqref{DSLgauge} is understood as the limit 
% keeping all orders in the genus expansion contributing in a finite manner 
that captures all orders in the genus expansion even
if $N_2$ goes to infinity \cite{Ling:2006up,Asano:2014vba}.
Since PWMM in this BPS sector is expressed by the matrix integral \eqref{matrix model},
the VEV of a quarter-BPS operator % , $\mathcal{O}$, for the matrix integral 
can be expanded in powers of $1/N_2$ as
\beq
	\label{'t Hooft expansion}
	\langle \mathcal{O}\rangle_M
	=
	\sum_{n=0}^\infty d_n(\lambda) N^{-n}_2,
\eeq
where 
$\lambda=g^2N_2$ % is the 't Hooft coupling. 
and
$\mathcal{O}$ is a quantity equivalent to the quarter-BPS operator, 
which is thus made of $M$.
As discussed in the previous section,
PWMM realizes the IIA LST on $S^5$ in the NS5-brane limit \eqref{DSLgauge}, 
which naturally suggests that the VEV is expanded with respect to
the effective LST coupling $\tilde g_s$.
Hence, %in the NS5-brane limit,
the expansion (\ref{'t Hooft expansion}) is rewritten as
\beq
	\label{eq:LST_expansion}
	\langle \mathcal{O}\rangle_M
	=
	C(\lambda)\sum_{n=0}^\infty c_n\,\tilde g_s^{n},
\eeq
where $C(\lambda)$ is an overall factor, which may be, in general, 
infinitely large in the limit.
The coefficients $c_n$ are supposed to be independent of $\lambda$ 
because any LST observable is considered as a finite function of $\tilde g_s$ 
up to the overall factor.
Therefore, if the NS5-brane limit \eqref{DSLgauge} successfully realizes IIA LST,
we will observe that the ratios $d_{n+1}(\lambda)/d_n(\lambda)$ have
the following $\lambda$-dependence:
% \cite{Ling:2006up, Asano:2014vba}
\beq
	\label{eq:ratio_coeff}
	\frac{d_{n+1}(\lambda)}{d_n(\lambda)}
	=
	\frac{c_{n+1}}{c_n}
	\lambda^{\frac{5}{8}}
	\exp\left[2^{\frac{3}{4}}\pi\frac{\lambda^{\frac{1}{4}}}{D}\right]
	,
\eeq
for each $n$.
This is a non-trivial prediction, which we examine in the next subsection.
% This gives a method of evaluating $\langle\mathcal{O}\rangle_M$ which contains the quantum effects
% in the IIA LST on $S^5$.

%%%%%%%%%%%%%%%%%%%%%%%%%%%%%%%%%%%%%%%%%%%%%%%%%%%%%
\subsection{Numerical analysis}
\label{Numerical analysis}

We simulate the quarter-BPS PWMM system \eqref{matrix model}
with $\Lambda=1$ % in \eqref{fuzzy sphere}
by using the hybrid Monte Carlo method.
In the viewpoint of the electrostatic system,
this case corresponds to the system consisting of one conducting disk at $z=d_1$ 
between two larger conducting disks at $z=0$ and $d_2$ 
in the background potential \eqref{PWMMpot_bg}.
While the conducting disk at $z=0$ has an infinitely large radius,
the other larger one at $z=d_2$ becomes infinitely large after the NS5-brane limit is taken.
% The one at $z=d_1$ is introduced in order to
% incorporate the effect from the matter in the geometry 
% into our numerical analysis.

The quarter-BPS operators we measure are constructed of the ``moduli'' matrix $M$,
which has the following structure in this case:
\beq
	M
	=
	(M_1\otimes \mathbf{1}_{D_{1}})\oplus
	(M_2\otimes \mathbf{1}_{D_{2}}) .
\eeq
% In this paper, we compute three types of quantities projected onto each block matrix:
% $\Tr M_s^2$, $\Tr M_s^4$ and $\Tr e^{-M_s}$ for $s=1$ and $2$.
In this paper, we compute $\Tr M_s^2$, 
which are projected traces onto each block matrix for $s=1$ and $2$.

In the following, we discuss two ways to see the validity of the NS5-brane limit.

\subsubsection*{Finiteness in the double scaling limit of PWMM}

\begin{figure}[t]
 \centering
 \includegraphics[width=100mm]{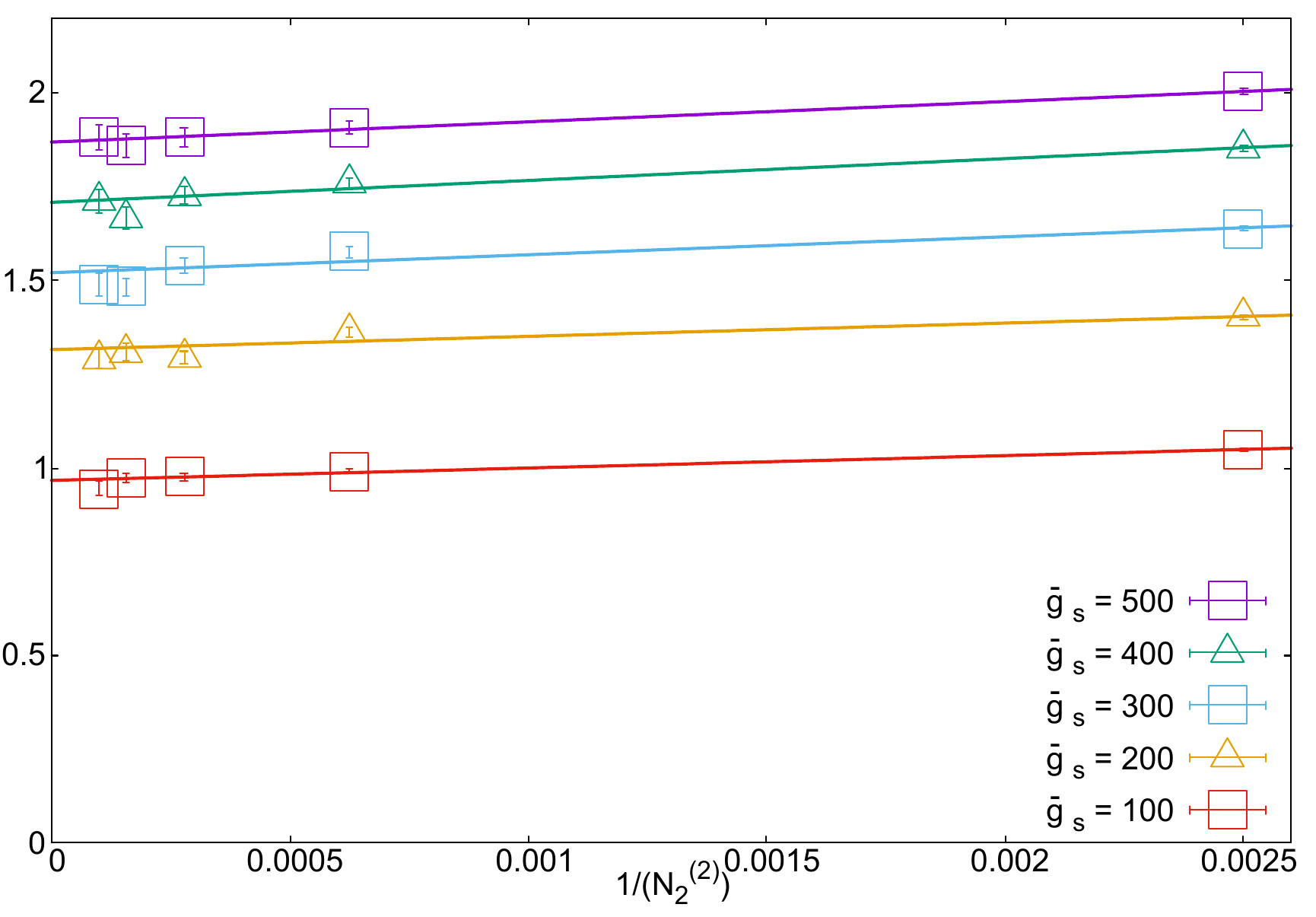}
 \caption
 {\small $N_2$-dependence of $\langle\Tr M^{\;2}_1\rangle_M$ for fixed values of $\tilde g_s$.
 Each set of data points for a fixed $\tilde g_s$ is fit by a linear function $a+b/N_2$, 
 where $a$ and $b$ are fitting parameters.}
 \label{double scaling limit of TrM^2}
\end{figure}

One way is to compute the quantities for $s=1$ at fixed $\tilde g_s$ 
and check whether they take finite values at large $N_2$ ($=N_2^{(2)}$).
Since $M_1$ corresponds to the middle conducting disk in the electrostatic picture
and therefore to a D2-brane flux in the supergravity picture,
the quantities for $s=1$ should describe a non-perturbative object in the IIA LST.
Thus, if the double scaling limit exists and reproduces the IIA LST,
such matrix-model quantities are expected to be finite in the limit in general.
% because, otherwise, the LST object cannot be described by them.
Therefore, $C(\lambda)$ in \eqref{eq:LST_expansion} should approach a finite constant
for $s=1$.
The plot of $\langle\Tr M_1^{\;2}\rangle_M$ is shown in
Fig.~\ref{double scaling limit of TrM^2} %for each value of $N_2$ %$N^{(2)}_2$
for $\tilde g_s=100$, 200, 300, 400 and 500
(see appendix \ref{sec:parameter_region for the NS5-brane limit}
% for our choice of $N_2^{(s)}$ and $N_5^{(s)}$).
for our choice of $N_2^{(s)}$ and $D_s$),
the values of which are chosen so that 
they satisfy $\lambda\gg D^{4}$ in \eqref{limit}.
One can see that they look convergent at each $\tilde{g}_s$
as $N_2$ tends to infinity.
This supports the validity of the NS5-brane limit.

\subsubsection*{Computation of $c_1/c_0$}
% Two-step fitting for $c_1/c_0$ computation

Another way is to compute quantities with the 't~Hooft coupling $\lambda$ fixed
and then read off the coefficients % in the $1/N_2$-expansion, 
$d_n(\lambda)$ in \eqref{'t Hooft expansion}
% (or $c_n$ through the double scaling relation \eqref{DSLgauge}) 
by fitting data by a polynomial function.
If the ratio $d_{n+1}(\lambda)/d_n(\lambda)$ behaves as in \eqref{eq:ratio_coeff} at large $\lambda$,
or equivalently, if the ratio $c_{n+1}/c_n$ approaches a constant as $\lambda$ goes to infinity,
finiteness of each term in the genus expansion is guaranteed even in the NS5-brane limit.
It means that the quantities have non-trivial dependence on $\tilde g_s$
in the computed parameter region, and therefore
that a system on NS5-branes is considered to be realized in the parameter region.

\begin{figure}[t]
	% \begin{minipage}{0.475\textwidth}
	% \centering
	% \includegraphics[width=\textwidth]{ratio_trM2_mod_vs_lambda.pdf}
	% \end{minipage}
	\begin{minipage}{0.475\textwidth }
	\centering
	\includegraphics[width=\textwidth]{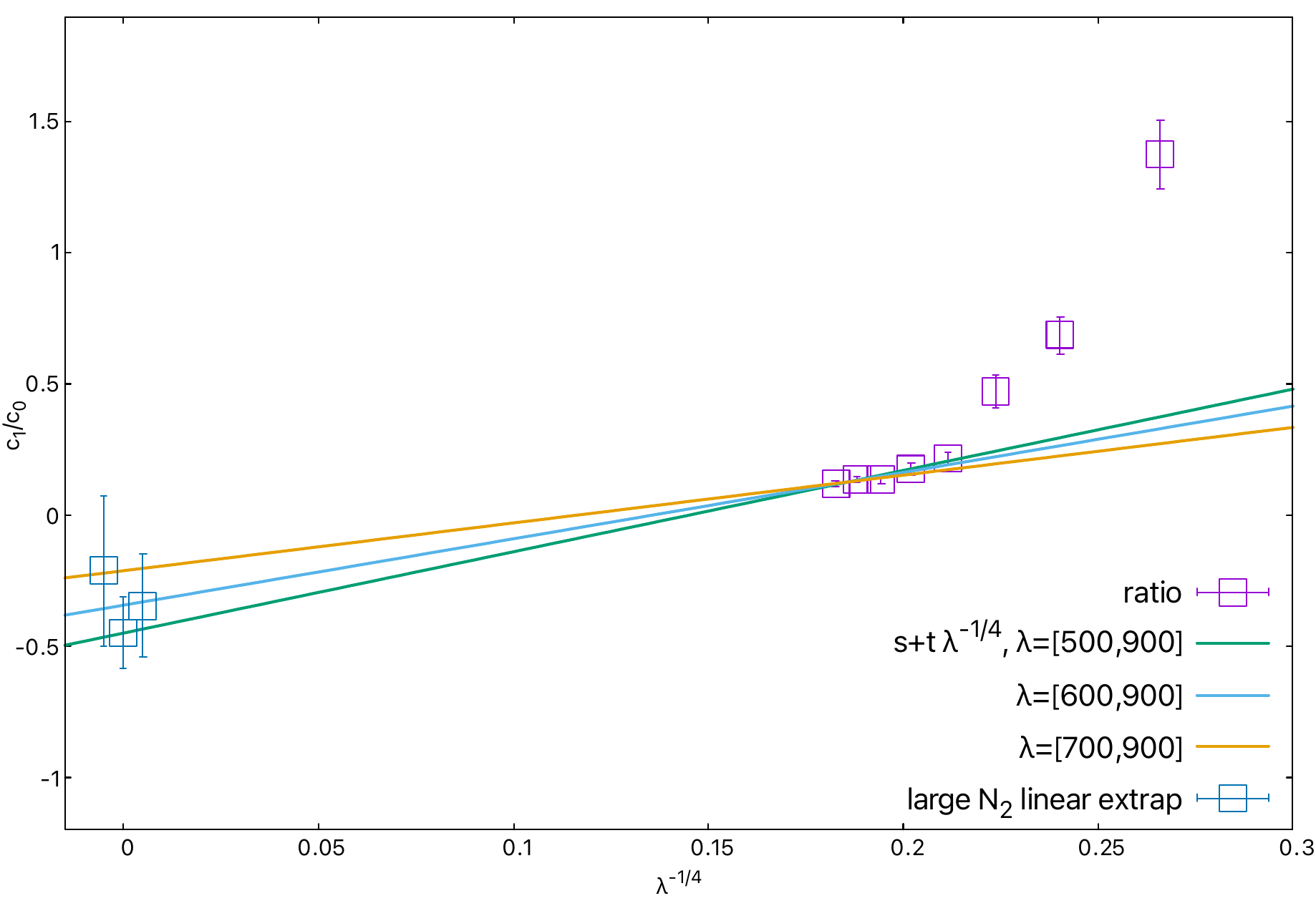}
	\end{minipage}
	\begin{minipage}{0.475\textwidth}
	\centering
	\includegraphics[width=\textwidth]{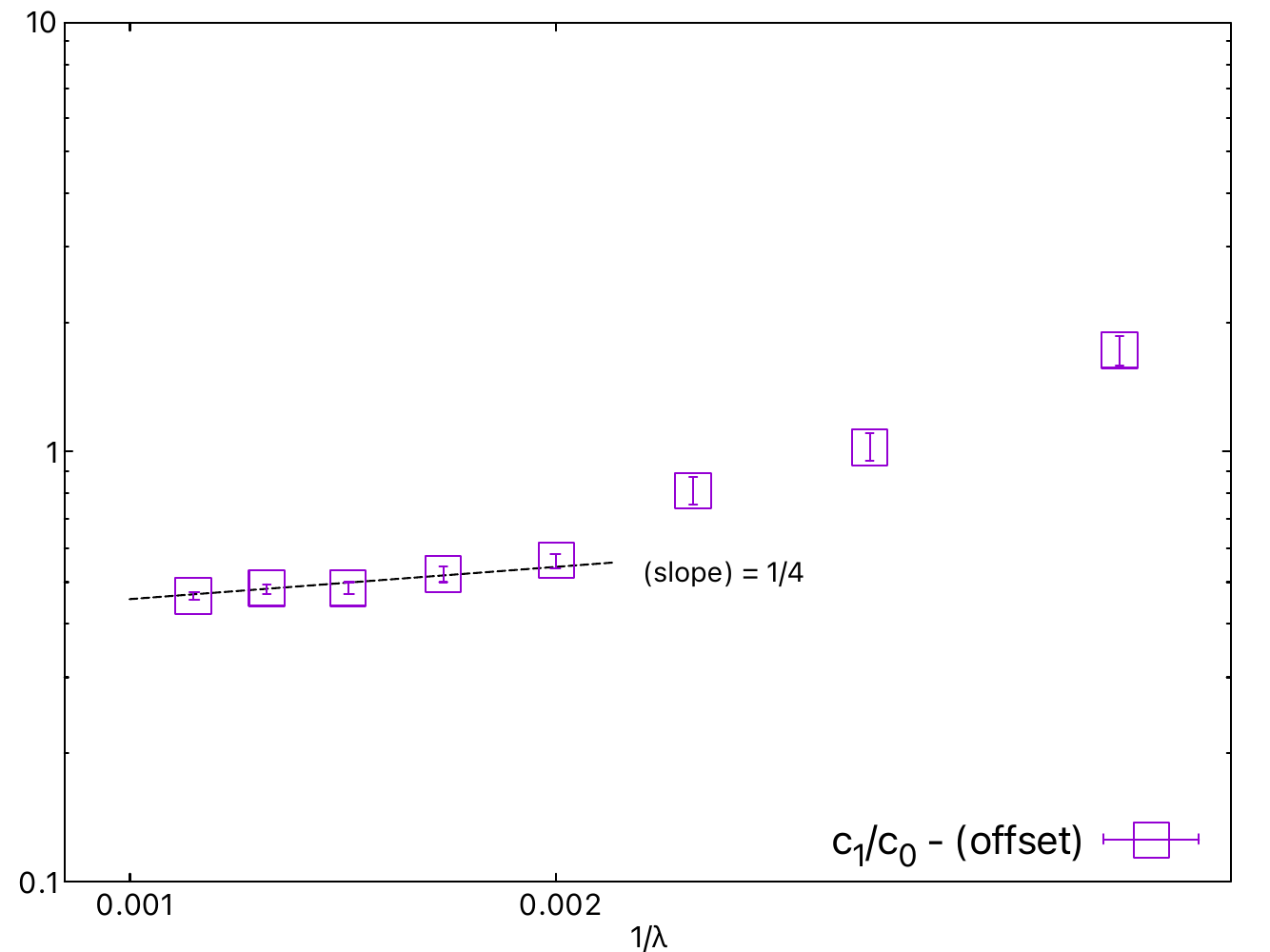}
	\end{minipage}
	\caption
	{\small The plots of the ratio $c_1/c_0$ for
	$\langle\Tr M^{\;2}_1\rangle_M$.
	[Left] The horizontal axis is $\lambda^{-1/4}$.
	The green, blue and orange lines are the fitted lines for different regions of $\lambda$, by $s + t \lambda^{-1/4}$,
        % and $s + t \lambda^{-1/4} + u \lambda^{-1/2}$, respectively 
        where $s$ and $t$ are fitting parameters.
	The blue squares just around $\lambda^{-1/4} = 0$ show the extrapolated value of $c_1/c_0$ by the 't~Hooft limit.
	[Right] The horizontal axis is $1/\lambda$.
	The black dashed line indicates a line with slope $1/4$.
	}
	\label{fig:ratio_trM2_vs_lambda}
\end{figure}

%Fig.~\ref{large N limit of TrM^2} shows a quadratic fit of $\langle\Tr M_1^{\;2}\rangle_M$ in powers of $1/N_2$ at $\lambda=200$.
The ratio $c_1/c_0$ at each $\lambda$ is computed 
by fitting of the numerically obtained expectation values. %of quantities
The expectation values are computed for several $N_2$, and hence, 
the large-$N_2$ extrapolation through the relation \eqref{eq:ratio_coeff} 
provides the coefficients, $d_0(\lambda)$ and $d_1(\lambda)$.
See appendix \ref{sec:parameter_region for the large-N limit}
% for our choice of $N_2^{(s)}$ and $N_5^{(s)}$ in this computation
for our choice of $N_2^{(s)}$ and $D_s$ in this computation
and appendix \ref{coefficients d0 d1} for $d_0(\lambda)$ and $d_1(\lambda)$ 
obtained by the fitting.
We plot $c_1/c_0$ for $\Tr M_1^{\;2}$
at $\lambda=200, 300, \cdots, 900$ in Fig.~\ref{fig:ratio_trM2_vs_lambda}.

We fit the obtained data of $c_1/c_0$ by a polynomial function of $\lambda^{-\frac{1}{4}}$
in the left panel in Fig.~\ref{fig:ratio_trM2_vs_lambda}.
This is motivated by the gauge/gravity duality,
which predicts that $\alpha'$-expansion corresponds to
expansion in powers of $\lambda^{-\frac{1}{4}}$ 
since $R_{S^5}^2/\alpha'=2\pi(8\lambda)^{\frac{1}{4}}$ in the NS5-brane limit\footnote{
$R_{S^5}$ is the $S^5$ radius at the spatial point corresponding to the edge of the highest conducting disk in the electrostatic picture.
} \cite{Asano:2014vba}.
The figure shows that
the linear fits in $\lambda^{-\frac{1}{4}}$ for data with $\lambda\ge 500$, 600 and 700
are consistent with each other
and that the ratio approaches $c_1/c_0\sim -0.3$.
Note % however 
that one cannot exclude the possibility that it converges to 0.

To numerically support the polynomial fitting in $\lambda^{-\frac{1}{4}}$,
we fit the log-log plot of $c_1/c_0$ against $1/\lambda$ 
by a linear function with slope $1/4$ 
in the right panel in Fig.~\ref{fig:ratio_trM2_vs_lambda}.
To properly detect the slope,
the ratio $c_1/c_0$ is subtracted by an offset $s=-0.34(20)$,
which is measured by the fitting $s+t\lambda^{-\frac{1}{4}}$ for $600\le\lambda\le 900$.
The good agreement %with the linear fitting by slope 1/4, 
seen in the figure reinforces the exponent $-1/4$ of $\lambda$ in the fitting polynomial 
and suggests that the correction at large $\lambda$ corresponds to
the $\alpha'$-correction to the NS5-brane system.

Another important check of the linear fitting is $\tilde g_s$-independence.
Even though the coefficients $c_i$ are supposed to be independent of $\tilde g_s$,
numerical results of $c_1/c_0$ may depend on $\tilde g_s$
because of numerical errors
which could be caused by the large values of $\tilde g_s$ in our parameter choice.
However, setting $\tilde g_s$ to the order of 1 
requires substantially larger $N_2$ than we used in the simulations.
Therefore, we extrapolate the data restricted to $\tilde g_s\le \tilde g_{s}^{\rm max}$,
with some constant $\tilde g_{s}^{\rm max}$,
and compare the extrapolated values of $c_1/c_0$ at large $\lambda$ 
with different $\tilde g_{s}^{\rm max}$.
We find that the extrapolation is not influenced drastically with $\tilde g_s^{\rm max}$.
See appendix~\ref{comparing_with_gs} for the detailed results.

%Firstly, we restrict the maximal value for $\tilde g_s$ to be used for the numerical extrapolation.
%In smaller $\lambda$ region, the matrix size $N_2$ to let $\tilde g_s$ be relatively small $(\lesssim O(10) )$ does not become so huge, and it enables us to generate many matrix configurations and obtain more precise expectation values of quantities by the hybrid Monte Carlo method.
%On the other hand, in larger $\lambda$ region, 

\begin{figure}[t]
	% \begin{minipage}{0.32\hsize}
	 % \begin{center}
	  \centering
	  \includegraphics[width=0.5\textwidth]{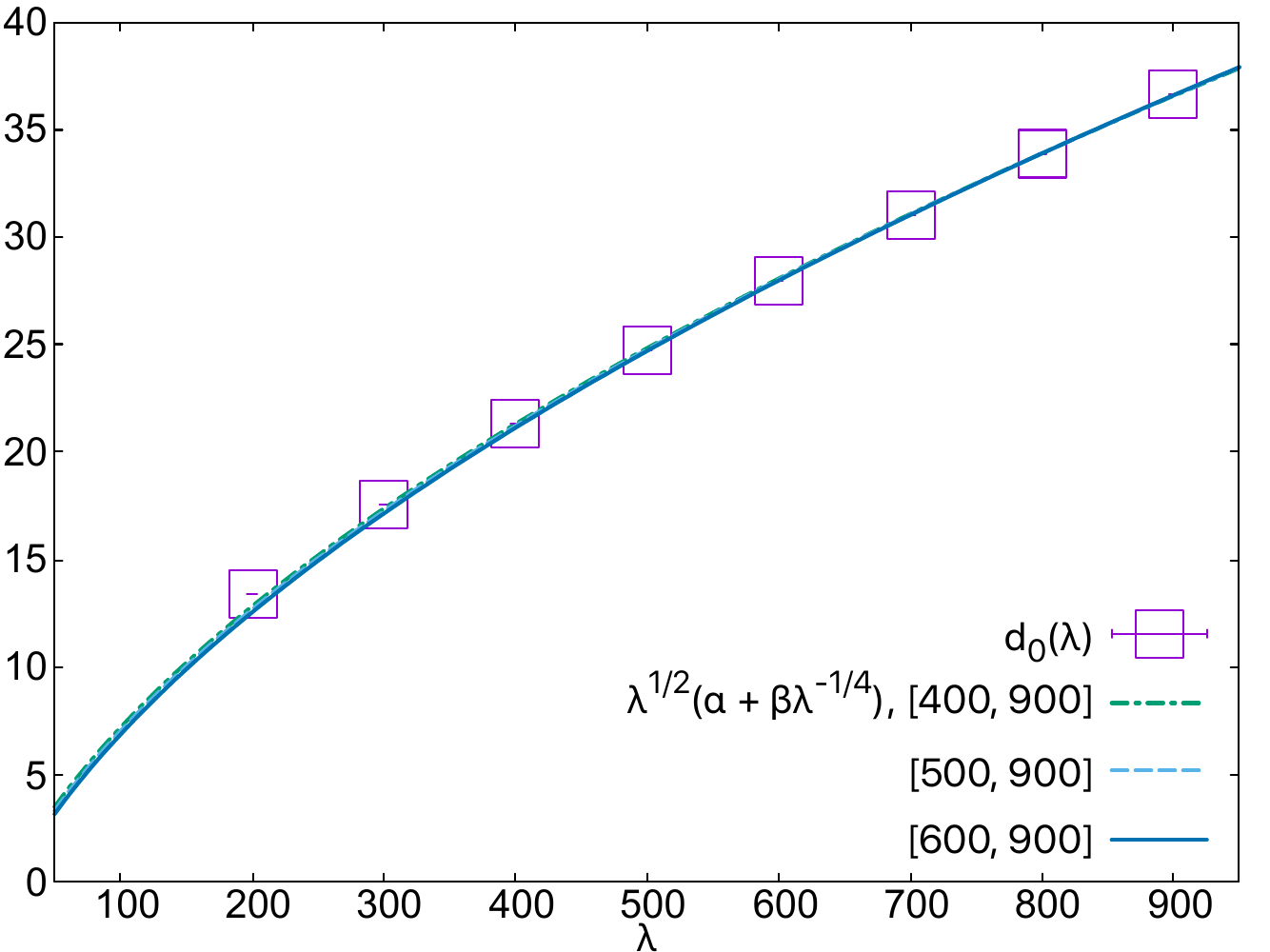}
	  % \includegraphics[width=55mm]{consistency_M2_1_v1.pdf}
	 % \end{center}
	% \end{minipage}
	% \begin{minipage}{0.32\hsize}
	%  \begin{center}
	%   \includegraphics[width=55mm]{consistency_M4_1_v1.pdf}
	%  \end{center}
	% \end{minipage}
	% \begin{minipage}{0.32\hsize}
	%  \begin{center}
	%   \includegraphics[width=55mm]{consistency_loop_9.pdf}
	%  \end{center}
	% \end{minipage}
	\caption
	{\small The plot of $d_0(\lambda)$ for $\langle\Tr M_2^{\;2}\rangle_M$
	% $\langle\Tr M_2^{\;4}\rangle_M$ and $\langle\Tr e^{-M_2}\rangle_M$.
	and some fits with different ranges.
	The curves show fitting results by
	$\lambda^{1/2}(\alpha+\beta\lambda^{-1/4})$ 
	with fitting parameters $\alpha$ and $\beta$.
	We see that it behaves consistent
	with the theoretical result
	at large $\lambda$
	.}
	\label{consistency}
\end{figure}

In order to verify that we simulate the system at large enough $\lambda$
to reproduce the physics on the NS5-branes,
we check the $\lambda$-dependence of $\langle\Tr M_2^{\;2}\rangle_M$.
It is shown in Ref.~\cite{Asano:2014vba} that 
this quantity grows as $\lambda^{\frac{1}{2}}$ at large $\lambda$,
i.e.~$C(\lambda)\sim\lambda^{\frac{1}{2}}$.
The next-to-leading order of the quantity is naturally assumed 
to be the order of $\lambda^{\frac{1}{4}}$, as discussed for the fitting of $c_1/c_0$.
We observe that numerical results of $\langle\Tr M_2^{\;2}\rangle_M$
agree 
with fits at large $\lambda$
by $\lambda^{1/2}(\alpha+\beta\lambda^{-1/4})$
with fitting parameters $\alpha$ and $\beta$, 
shown in Fig.~\ref{consistency}.
Our data is consistent with the theoretically expected behavior;
hence this supports the validity of our parameter choice.

\begin{figure}[htbp]
 \centering
 \includegraphics[scale=0.4]{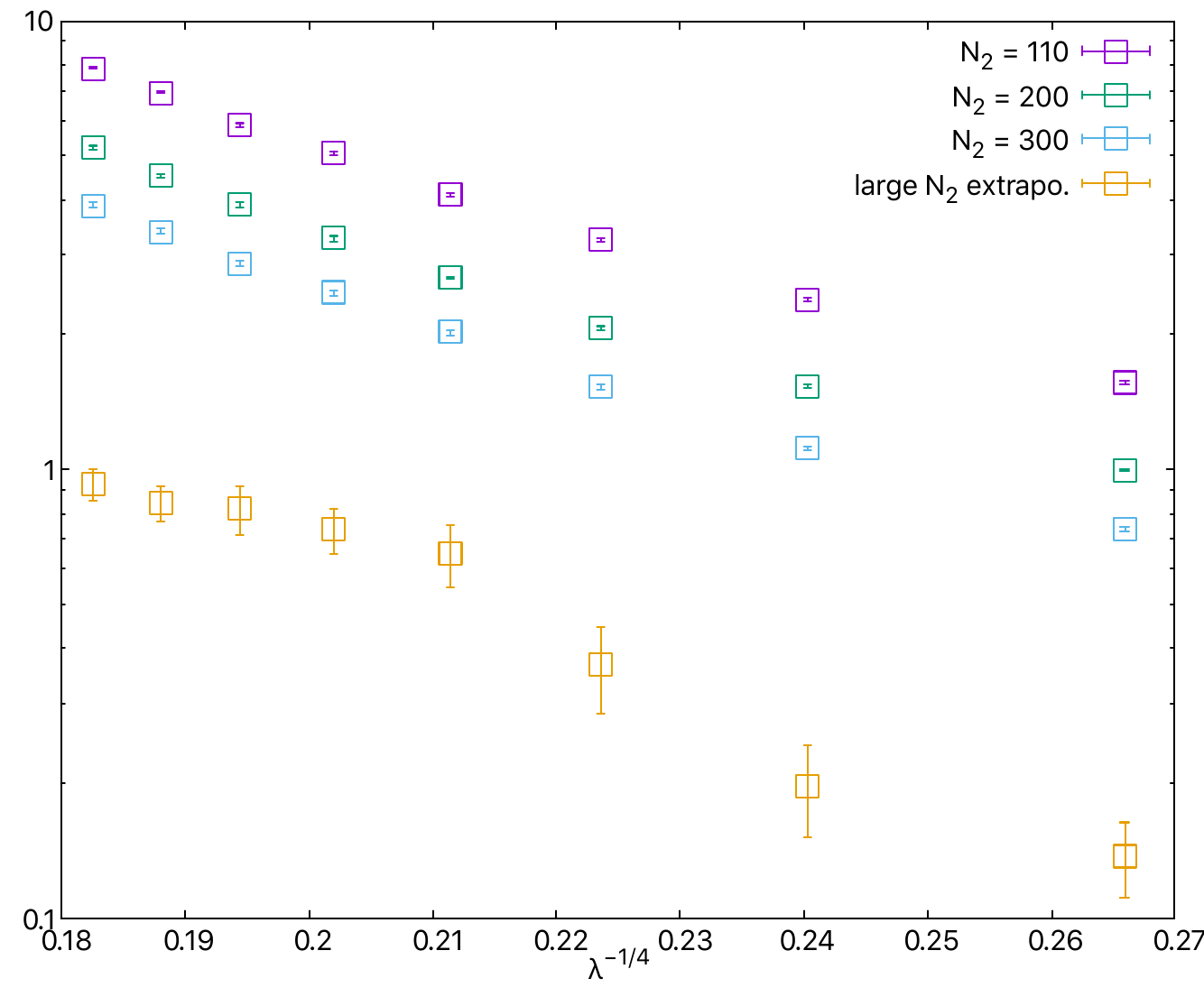}
 \caption{\small The plot of $\langle\Tr M_1^{\; 2}\rangle_M$ against $\lambda^{-\frac{1}{4}}$.
 The purple, green and blue points are numerical results
 with $N_2=100$, 200 and 300, respectively.
 The orange points represent extrapolated values at large $N_2$.
 Around $\lambda^{-\frac{1}{4}}=0.22$,
 the behavior of the extrapolated points changes rather drastically,
 which may suggest there is a phase transition.}
 \label{fig:trM2_vs_lambda_N2-dep}
\end{figure}

Even though the above arguments support the linear fitting 
in Fig.~\ref{fig:ratio_trM2_vs_lambda} to some extent,
the deviation from the linear fit does not look like a quadratic correction.
% as seen in Fig.~\ref{fig:ratio_trM2_vs_lambda}.
However, since the 5-brane physics is not well understood yet,
the exact correction to the linear term, 
which should correspond to higher $\alpha'$-corrections,
is simply not known.
Thus, it is possible that some higher-order terms are accidentally large or 
that the fitting function is not even a polynomial but in a more complicated form.
% In addition, it is also possible that there is a phase transition around $\lambda=450$.
In addition, it is also a possible scenario that there is a phase transition around $\lambda=450$, which may be caused by the large-$N_2$ limit.
(See Fig.~\ref{fig:trM2_vs_lambda_N2-dep}.)

%%%%%%%%%%%%%%%%%%%%%%%%%%%%%%%%%%%%%%%%%%%%%%%%%%%%%%%%%%%%%
\section{Summary and discussion}
%%%%%%%%%%%%%%%%%%%%%%%%%%%%%%%%%%%%%%%%%%%%%%%%%%%%%%%%%%%%%

We investigated a novel realization of the IIA LST on $R\times S^5$ by PWMM in a double scaling limit, 
which is called the NS5-brane limit throughout this paper.
The IIA LST on $R\times S^5$ is known to have many discrete vacua and the theory around each vacuum is
conjectured to be realized by PWMM in the limit.
For the trivial vacuum, the form of the NS5-brane limit 
was obtained in \cite{Ling:2006up},
where the limit was first obtained in the gravity side and translated into the field-theory language 
based on some reasonable arguments.

In this paper, we obtained the NS5-brane limit for general vacua.
We first found the NS5-brane limit on the gravity side in terms of 
the electrostatic description. Then, by making use of the localization method,
we showed the equivalence between 
the effective action of a quarter-BPS sector in 
PWMM and the action for the electrostatic system.
This equivalence made it possible to directly identify the parameters on both sides.
% and determine the form of the NS5-brane limit in PWMM.
Using this relation, we finally expressed the form of the NS5-brane limit 
in terms of the parameters in PWMM.

The use of the localization computation also yields a byproduct that 
one can see the direct equivalence between the quarter-BPS sector in PWMM 
and the electrostatic system on the gravity side.
This equivalence assures that the NS5-brane limit, which we found in 
the gravity side, also exists in the quarter-BPS sector in PWMM.
However, since we took the planar limit of PWMM in 
deriving this equivalence, our analytical proof on the presence of the NS5-brane limit in PWMM
is also limited to this sector. 
Note that our numerical results in section \ref{sec:numerical_results} presented
evidence of the NS5-brane limit in more general sector beyond the planar limit.

%In obtaining this, we made use of the relations among geometries, electrostatic systems and 
%vacua of PWMM and LST. 
%% We first showed that the effective action describing a quarter BPS sector of PWMM can be written as that for
%% the corresponding electrostatic system, the potential of which is directly related to its dual geometry.
%% Then, we found the double scaling limit in which the effective action for PWMM
%% correctly reproduces that for the electrostatic system for LST around a general vacuum.
%% In the gauge theoretic language, this indicates that
%% IIA LST on $R\times S^5$ would be obtained by the double scaling limit of PWMM.

%% This limit corresponds, in the language of the electrostatic system, 
%% to sending the radius of the disk at the highest position 
%% to infinity with the strength of the background potential tuned so that that for the NS5 brane 
%% geometry is realized. 
%% We found that the large radius limit of the disk is responsible for the infinitely many mirror images
%% of finite disks, which

The IIA LST around a trivial vacuum can be also obtained by different double scaling limits of the other $SU(2|4)$ symmetric theories, 
$\mathcal{N}=8$ SYM on $R\times S^2$ and 
$\mathcal{N}=4$ SYM on $R\times S^3/Z_k$ \cite{Ling:2006xi}.
By applying the procedure used in this paper,
it would be possible to show the existence of the double scaling limit 
to obtain the IIA LST around a general vacuum from these gauge theories.

%While we only focused on a vacuum of LST, it will be possible to extend the 
%argument in this paper into the cases with some insertions of the BPS 
%operators. 
%In PWMM, the vev of BPS operators made of $\phi(\tau)$ can be evaluated 
%by the same effective action. 
%As long as the dimensions of the operators are small enough, 
%not large compared to $N$,
%one can take the same double scaling limit for such operators, 
%which result in some BPS operators of LST.
%It will be nice if one can clarify physical meaning of those BPS operators 
%in LST 
%and utilize them to reveal unknown features of LST.

In the numerical computation, 
we obtained results consistent with the existence of the double scaling limit.
% we found that it supported the existence of the double scaling limit.
We also found that the coefficients in the $\tilde g_s$ expansion of 
$\Tr M_1^{\; 2}$, which is associated with the electric charge density 
(or some sort of D2-brane charge density in the supergravity picture),
satisfied $c_1/c_0\sim -0.3$. %-0.4
% and that quantities associated with the electric charge density 
% (or some sort of D2-brane charge density in the supergravity picture)
% was independent of $\tilde g_s$.

Our result is the first computation 
to obtain $\tilde g_s$ correction in LST using PWMM, 
which can be interpreted as quantum loop correction
in the Lin-Maldacena geometry in the limit \eqref{tHooft limit}, 
to the best of our knowledge.
However, the linear fitting of $c_1/c_0$ was not quite satisfactory.
This should be due to the lack of knowledge of the 5-brane physics.
In this regard, it would be interesting to clarify 
whether there is a phase transition in the large-$N$ limit,
inferred in Fig.~\ref{fig:trM2_vs_lambda_N2-dep}.

Since the error is large, the ratio $c_1/c_0$ for $\Tr M_1^{\; 2}$ was
consistent with 0,
which means $\Tr M_1^{\; 2}$ would be independent of $\tilde g_s$.
If this is the case,
the $\tilde g_s$-independence of the quantity 
would suggest that there is a non-renormalizable theorem,
or it is equally possible that 
only the next-to-leading correction in $\tilde g_s$ is absent 
while higher-order corrections are present.
Unfortunately, it is not feasible to determine 
the higher-order terms to high precision by the current numerical method,
or at least, in the parameter region of our choice.

The relation of the obtained double scaling limit with the M-theoretic viewpoint 
is also interesting.
In \cite{Asano:2017xiy,Asano:2017nxw},
it was revealed that, in the strong coupling limit of PWMM, 
the low energy modes of the $SO(6)$ matrices form the $S^5$ geometry 
with the correct radius predicted in \cite{Maldacena:2002rb}.
Even though its interpretation of the emerged geometries 
is slightly different from the gauge/gravity,
they should be related in some way.

%%%%%%%%%%%%%%%%%%%%%%%%%%%%%%%%%%%%%%%%%%%%%%%%%%%%%%%%%%%%%
\section*{Acknowledgments}
%%%%%%%%%%%%%%%%%%%%%%%%%%%%%%%%%%%%%%%%%%%%%%%%%%%%%%%%%%%%%
The work of G.~I.~was supported by JSPS KAKENHI Grant Number JP~19K03818.
H.~W.~was supported in part by JSPS KAKENHI Grant Number JP~21J13014.

\appendix
\section{Solution in the NS5-brane limit with $\Lambda=1$}\label{apx:solution}
The solution to \eqref{f0 eq} was obtained in \cite{Ling:2006up}.
See \cite{Asano:2014vba} for the gauge theoretic viewpoint.
In \cite{Ling:2006up}, 
the equation \eqref{f0 eq} 
was solved by imposing the constraint that 
the derivative of $f^{(0)}$ 
is vanishing at the edges of the support of $f^{(0)}$.
This comes from the finiteness of the corresponding gravity solution.
Note that from the gauge theoretic viewpoint, 
the meaning of this condition is not clear 
and it seems to be unnatural to impose the condition by hand 
on the corresponding matrix integral discussed in \cite{Asano:2014vba}.
However, it turns out that, on the gauge theory side, 
the solution with the condition imposed is chosen by 
the least action principle.
In this appendix, we derive the solution from the action principle, 
not imposing the condition.

Let us start with the action of the electrostatic system,
%shown in section \ref{NS5 general},
%which produces \eqref{f0 eq} as its saddle point equation. 
which is given by \eqref{ses} with $\Lambda=0$:
\begin{align}
S_{eff}
 &=\frac{32}{\pi^4}\Bigg[
 2V_0d \int_{-R}^{R} du \ u^2 f^{(0)}(u)
 +\frac{1}{2}  \int du\, f^{(0)}(u)^2
 \nonumber \\
 &\quad
 -\frac{1}{2\pi} \int du du' \left[
 \frac{2d}{(2d)^2+(u-u')^2}
 \right] f^{(0)}(u)f^{(0)}(u') 
 \nonumber \\
 &\quad
 -(C^{(0)}+\frac{2}{3}V_0d^3)\left(\int du\, f^{(0)}(u)-\pi Q\right)
 \Bigg].
 \label{S_eff Lambda=1}
\end{align}
The saddle point equation of $f^{(0)}$ is given by
\eqref{f0 eq}.
If we take the $d/R\to 0$ limit in \eqref{f0 eq},
we obtain 
\begin{align}
 \omega^{(0)\prime}(u-i0)+\omega^{(0)\prime}(u+i0)
 &=\mathcal{C}-2\pi V_0 u^2,
 \label{saddle pt eq in NS5 lim}
\end{align}
where $\mathcal{C}=\frac{\pi}{d}(C^{(0)}+\frac{2}{3}V_0d^3)$ and $\omega^{(0)}(z)$ is the resolvent:
\begin{align}
 \omega^{(0)}(z)=\int_{-R}^{R} du\frac{f^{(0)}(u)}{z-u}.
\end{align}
If we integrate \eqref{saddle pt eq in NS5 lim} once, 
we obtain an equation which is   
equivalent to the equation of motion of the 
one-matrix model with a quartic interaction. 
Thus, we find that the solution is given as the same form as the 
quartic matrix model:
\begin{align}
 \omega^{(0)} (z)
 =\frac{1}{2}\left( \mathcal{C} z -\frac{2\pi V_0}{3}z^3 \right) 
 -\frac{1}{2} \Bigg( 
 \mathcal{C} -\frac{2\pi V_0R^2}{3}\left( \frac{1}{2}+\frac{z^2}{R^2}\right)
 \Bigg) \left( z^2-R^2 \right)^{\frac{1}{2}}.
 \label{resolvent0}
\end{align}
Furthermore the normalization 
\eqref{QR} gives another relation as
\begin{align}
 \mathcal{C}
 =\pi\frac{V_0R^{4}+8Q}{2R^2}.
 \label{cC}
\end{align}
By eliminating ${\mathcal C}$ using this relation, 
we obtain
\begin{align}
 f^{(0)}(u)
 =\frac{V_0}{3} \Bigg( 
 \frac{V_0R^{4}+24Q}{4R^2V_0}
 -u^2
 \Bigg) \left( R^2-u^2 \right)^{\frac{1}{2}}.
 \label{f0 with R undetermined}
\end{align}
Note that the semi-positivity of $f^{(0)}(u)$ imposes the condition that
$R\leq (8Q/V_0)^{\frac{1}{4}}$.

%The remaining task is determining $R$.
%However, 
%the normalization condition, 
%which usually determines $R$ as in the Gaussian matrix model,
%is already used to fix $\mathcal{C}$.
%Also, the condition $f^{(0)}(\pm R)=0$ clearly imposes nothing on $R$.
%The point is that the solution of $f^{(0)}(u)$ minimizes the 
%effective action\footnote{
%Note that the effective action \eqref{S_eff Lambda=1} is 
%the one on the gauge theory side.
%From the viewpoint of the corresponding electrostatic system,
%it is energy that is minimized by the solution.
%}.
Then, we use the action principle. By substituting
\eqref{f0 with R undetermined} into \eqref{S_eff Lambda=1},
we find that the $R$-dependence of the effective action is given by
\begin{align}
 S_{eff}=
 \frac{d}{3\pi^3R^2}\left(
 192 Q^2 + 48\pi V_0Q R^2 - \pi^2V_0^2R^6
 \right) .
\end{align}
This has an extremum point at
\begin{align}
 R=\left( \frac{8Q}{V_0} \right)^{\frac{1}{4}}.
\label{R at extremum}
\end{align}
Under the condition that $f^{(0)}(u)$ is semi-positive everywhere,
we can find that this point is also the minimum of the action.
Now, by substituting (\ref{R at extremum}), we obtain
\begin{align}
 f^{(0)}(u)
 =\frac{V_0R^3}{3} \left\{ 1-\left(\frac{u}{R}\right)^2 \right\}^{\frac{3}{2}}
 =\frac{V_0}{3} \left( \sqrt{\frac{8Q}{V_0}}-u^2 \right)^{\frac{3}{2}}.
\label{critical f0}
\end{align}
Thus, the solution with $\frac{df^{(0)}}{du}(\pm R)=0$ is 
indeed obtained through 
the least action principle.

The same procedure %to determine $x_m$ (or $\bar\mu$) 
can be also applied to 
calculating \eqref{rho critical}.
The procedure is as follows.
Firstly, in the NS5-brane limit, 
the leading behavior of $\rho(x)$ is governed only by self-interaction
since the effect from the eigenvalues for $s=1,\cdots,\Lambda$ is 
on the order of $\frac{Q_s}{Q}$ and hence suppressed.
Then, by plugging \eqref{rho not critical} and \eqref{bar_mu} into 
the effective action \eqref{effective action} 
and neglecting the interaction terms involving 
$\rho^{(s)}(x)$ for $s=1,\cdots,\Lambda$,
one obtains the expression of the action written by $x_m$
in the leading of $D/x_m$.
% as $S_{eff}\simeq D\left( \frac{2 N_2^2}{x_m^2} + \frac{N_2 x_m^2}{2 g^2} - \frac{x_m^6}{96 g^4} \right) $.
Finally, one finds that its minimum point is given by \eqref{xm^4}.

\section{Derivation of \eqref{PWMMpotential2parts}}\label{apx:potential_recursion}

In this appendix, we derive \eqref{PWMMpotential2parts} from \eqref{gIntEq}.

%% In this section, 
%% we discuss a recurrence relation generated by \eqref{gIntEq},
%% which is the root of the infinitely many mirror images in \eqref{NS5potential},
%% and then derive \eqref{PWMMpotential2parts}.

%% To begin with, 
%% in order to discuss the convergence of equations properly,
%% let us rescale and redefine the charge densities as
%% \begin{align}
%%  \bar f_{s}(x):=8\frac{f_{s}(R x)}{V_0R^3},
%%  \qquad
%%  \bar g(x):=8\frac{g(Rx)}{V_0R^3},
%% \end{align}
%% and constants as
%% $\bar C_{s}:=8C_{s}/(V_0R^3)$ and
%% $\bar C^{(0)}:=8C^{(0)}/(V_0R^3)$
%% for $s=1,\cdots,\Lambda+1$.
%% Then, \eqref{gIntEq} is rewritten as a finite equation:
%% \begin{align}
%%  \bar g(x)-K_{\Lambda+1}(2d)\bar g(x)
%%  -\sum_{t=1}^{\Lambda}(K_t(d+d_t)-K_t(d-d_t))\bar f_t(x)
%%  % \nonumber \\
%%  =\bar C_{\Lambda+1}-\bar C^{(0)}
%%  .
%%  \label{bar_gIntEq}
%% \end{align}

We first show that the right-hand side of \eqref{gIntEq}, $C_{\Lambda+1}- C^{(0)}$, vanishes in the NS5-brane limit
\eqref{DSLgrav}.
For this purpose, we integrate \eqref{gIntEq} over $u$ from $-R$ to $R$:
\begin{align}
 C_{\Lambda+1}- C^{(0)}
 &=\frac{2d}{\pi R}
 \int_{-R}^{R}du\, \frac{g(u)}{R^2-u^2}
 +\sum_{s=1}^{\Lambda}\frac{2d_s}{\pi R}
 \int_{-R_s}^{R_s}du\, \frac{f_s(u)}{R^2-u^2}
 +\cdots
 % \nonumber \\
 % &=\frac{2d}{\pi R}
 % \int_{-1}^{1}dx'\, \frac{\bar g(x')}{1-x'^2}
 % +\sum_{t=1}^{\Lambda}\frac{2d_t}{\pi R}\frac{8Q_t}{V_0R^4}
 % +\cdots
 ,
 \label{C-C0}
\end{align}
where we have used
\begin{align}
 \int_{-R}^{R} du\, \frac{\delta}{\delta^2+(u-u')^2}
 &=\pi 
 -\frac{2R\delta}{R^2-u'^2}
 +\mathcal{O}((\delta/R)^3)
 .
 % \nonumber \\
 % &=\pi 
 % -2\left( 1+x'^2 \right) \frac{\delta}{R}
 % +O(x'^4,(\delta/R)^3)
\end{align}
As noted below \eqref{f Lambda+1}, $g(u)$ and $f_s(u)$ are on the order of $Q_s$ and $g(\pm R)=0$.
Using these conditions, one can show that the right-hand side of \eqref{C-C0} vanishes in the NS5-brane limit.
Thus, in the NS5-brane limit,  $C_{\Lambda+1}- C^{(0)}\to 0$.

Next, we introduce a function,
\begin{align}
 % G(\delta,r,z)
 % &:=\frac{1}{\pi}\int_{-R}^{R}du\, 
 % % \left(
 % \frac{1}{\sqrt{(z-\delta+iu)^2+r^2}}
 % % -\frac{1}{\sqrt{(z+\delta+iu)^2+r^2}}
 % % \right) 
 % g(u)
 % \nonumber \\
 G(\delta,r,z)
 &:=\frac{1}{\pi}\int_{-R}^{R}du\, 
 \frac{1}{\sqrt{(z-\delta+iu)^2+r^2}}
 g(u)
 .
 \label{pot for g}
\end{align}
In terms of this function, the potential produced by $g(u)$ 
is just given by $G(d,r,z)-G(-d,r,z)$.
Plugging \eqref{gIntEq} into \eqref{pot for g} with $\delta=\pm d$ and 
taking the NS5-brane limit \eqref{DSLgrav}, 
we obtain
\begin{align}
  G(\pm d,r,z)
 &= G(\pm 3d,r,z)
 % \nonumber \\
 % &\quad 
 +\sum_{s=1}^{\Lambda}
 \frac{1}{\pi}\int_{-R_s}^{R_s}du\, 
 \Bigg(
 \frac{1}{\sqrt{(z\mp (2d+d_s)+iu)^2+r^2}}
 % -\frac{1}{\sqrt{(z+2d+d_t+iu)^2+r^2}}
 \nonumber \\
 &\hspace{65mm}
 -\frac{1}{\sqrt{(z\mp (2d-d_s)+iu)^2+r^2}}
 % +\frac{1}{\sqrt{(z+2d-d_t+iu)^2+r^2}}
 \Bigg)  f_s(u),
 \label{recursion}
\end{align}
where we have % neglected the terms of $\mathcal O(\frac{Q_s^2}{Q^2})$, 
% which appear since \eqref{gIntEq} holds only for $|u|<R'$, and 
used $C_{\Lambda+1}- C^{(0)}\to 0$ and
the following relation that holds as $R\to\infty$:
\begin{align}
 % &\lim_{R\rightarrow \infty}
 &\int_{-1}^{1}du\, 
 \frac{1}{\sqrt{(z+iu)^2+r^2}}  K_s(\delta) f_s(u) 
 \to \int_{-R_s}^{R_s}du'\, \frac{1}{\sqrt{(z+\mathrm{sgn}(z)\delta+iu')^2+r^2}} f_s(u').
\end{align}
Repeating this procedure $\nu$ times, we arrive at
\begin{align}
  G(\pm d,r,z)
 &= G(\pm (2\nu+1)d,r,z)
 \nonumber \\
 &\qquad 
 \pm \sum_{n=1}^{\nu}\sum_{s=1}^{\Lambda}
 \frac{1}{\pi}\int_{-R_s}^{R_s}du\, 
 \Bigg(\frac{1}{\sqrt{(z-d_s\mp 2nd+iu)^2+r^2}}
 \nonumber \\
 &\qquad\qquad\qquad\qquad\qquad\qquad
 -\frac{1}{\sqrt{(z+d_s\mp 2nd+iu)^2+r^2}}\Bigg)  f_s(u).
 \label{recursion_nu}
\end{align}
Note that %, using the mean-value theorem, 
one can show that $G(\delta,r,z)$ is finite even in the NS5-brane limit
and that $G(\delta,r,z)$ is on the order of $1/\delta$ for sufficiently large $\delta$.
Thus, the first term on the right-hand side turns out to vanish as $\nu\to\infty$.
% We now let $\nu$ go to infinity
% before taking the $R\to\infty$ limit completely.
% Then the first term in the right-hand side turns out to vanish % because of \eqref{g}.
% as $\nu\to \infty$
% because $G(\pm (2\nu+1)d,r,z)$ is finite when $R$ is fixed to a finite value.
From the $\nu\to\infty$ limit of this expression,
we finally obtain relation \eqref{PWMMpotential2parts}.

\section{Electrostatic action in the NS5-brane limit}\label{apx:elextrostatic_action}
In this appendix, we derive the action of the electrostatic system for LST
by starting from the action for PWMM \eqref{ses} and then
taking the NS5-brane limit.

A tricky term in \eqref{ses} involved in the derivation is 
the term of $C_{\Lambda+1}\sigma_{\Lambda+1}(r)$.
This can be rewritten as
\begin{align*}
 2C_{\Lambda+1}\int_0^\infty 2\pi rdr\, \sigma_{\Lambda+1}(r)
 =-\frac{1}{4\pi}C_{\Lambda+1}
 \int_{-\infty}^{\infty} dz \int_0^\infty 2\pi rdr\, 
 \frac{z}{|z|}\bigtriangleup V_{\Lambda+1}(r,z) .
\end{align*}
Then, this $V_{\Lambda+1}$ can be divided into $V^{(0)}$ and $V_{s,n}$ % obtain the following, 
by using the equation \eqref{PWMMpotential2}; %acted by $z/|z|\bigtriangleup$:
% holds in the 
% large-$R$ limit:
% \begin{align}
%  \frac{z}{|z|}\bigtriangleup  V_{\Lambda+1}(r,z)
%  &=\frac{z}{d}\bigtriangleup  V^{(0)}(r,z)+
%  \sum_{s=1}^{\Lambda}
%  \sum_{\substack{n=-\infty \\ \neq 0}}^{\infty}
%  \frac{z}{|z|}
%  \bigtriangleup V_{s,n}(r,z).
%  \label{z nabla V}
% \end{align}
% One can derive this
% by noting that the equation \eqref{PWMMpotential2parts} for each sign holds independently
% and that the charge densities satisfy the following relations,
% \begin{align}
%  &\frac{1}{\pi}\bigtriangleup\int_{-R}^{R} du 
%  \frac{1}{\sqrt{(z-d+iu)^2+r^2}} f^{(0)}(u)
%  =-4\pi \sigma^{(0)}(r)\delta(z-d),
%  \nonumber \\
%  &\frac{1}{\pi}\bigtriangleup\int_{-R_s}^{R_s} du 
%  \frac{1}{\sqrt{(z-d_s-2nd+iu)^2+r^2}} f_s(u)
%  =-4\pi \sigma_s(r)\delta(z-d_s-2nd).
%  \nonumber
% \end{align}
hence it becomes
\begin{align}
 2C_{\Lambda+1}
 \int_0^\infty 2\pi rdr\, 
 \sigma_{\Lambda+1}(r)
 &=2C_{\Lambda+1}
 \int_0^\infty 2\pi rdr\, 
 \sigma^{(0)}(r)
 \label{Csigma_Lambda+1}
\end{align}
because % $\frac{z}{|z|}\bigtriangleup V_{s,n}(r,z)$ is
\begin{align}
 &\frac{z}{|z|}\bigtriangleup V_{s,n}(r,z)
 =-4\pi \sigma_s(r)(\delta(z-d_s-2nd)-\delta(z+d_s-2nd))
 \label{LaplacianVsn}
\end{align}
holds for $|n|\geq 1$ and its integration over $z$ vanishes.
However, in order to express the action for LST by $V_{s,n}(r,z)$ in a neat way,
we rewrite \eqref{Csigma_Lambda+1} as
\begin{align}
 2C_{\Lambda+1}
 \int_0^\infty 2\pi rdr\, 
 \sigma_{\Lambda+1}(r)
 &=2C_{\Lambda+1}
 \int_0^\infty 2\pi rdr\, 
 \sigma^{(0)}(r)
 \nonumber \\
 &\quad
 -\frac{1}{4\pi}
 \int_{-\infty}^{\infty} dz \int_0^\infty 2\pi rdr\, 
 \sum_{s=1}^{\Lambda}
 \sum_{\substack{n=-\infty \\ \neq 0}}^{\infty}
 C_{s}\frac{z}{|z|}
 \bigtriangleup V_{s,n}(r,z) ,
 \label{Csigma_Lambda+1_fin}
\end{align}
by utilizing the fact 
that the integration of \eqref{LaplacianVsn} over $z$
vanishes for arbitrary $s$ when $|n|\geq 1$.

By substituting \eqref{eom_for_pot}, \eqref{PWMMpotential3} 
and \eqref{Csigma_Lambda+1_fin} into \eqref{ses}
and gathering the terms relevant for $V_{s,n}(r,z)$ (or $f_{s}(u)$),
we now obtain, up to surface terms,
\begin{align}
 S_{es}
 &=-\frac{1}{4} \int_{-\infty}^{\infty}dz  \int_{0}^{\infty}rdr
 \Bigg[
 -\sum_{s,n}\sum_{t,m} V_{s,n}(r,z)\bigtriangleup  V_{t,m}(r,z)
 \nonumber\label{periodic S_eff} \\
 &\qquad
 -2\sum_{s,n}\left( 
 V_0(r^2z-\frac{2}{3}z^3)+ V^{(0)}(r,z)-C_s\frac{z}{|z|}
 \right) \bigtriangleup  V_{s,n}(r,z)
 \Bigg]
 \nonumber \\
 &\qquad
 -2\sum_{s=1}^{\Lambda}C_sQ_{s}
 +S_{es,\Lambda+1},
\end{align}
where $S_{es,\Lambda+1}$ is an irrelevant term:
\begin{align*}
 S_{es,\Lambda+1}
 &=-\frac{1}{4} \int_{-\infty}^{\infty}dz  \int_{0}^{\infty}rdr
 \Bigg[
 -\left( 
 V_0(r^2z-\frac{2}{3}z^3)+ V^{(0)}(r,z)
 \right) \bigtriangleup  V^{(0)}(r,z) \Bigg]
 \nonumber \\
 &\qquad
 +2C_{\Lambda+1}
 \left( \int_0^\infty 2\pi rdr\, 
 \sigma^{(0)}(r) -Q \right) .
\end{align*}
By putting $C_s=\tilde C_s+C^{(0)}\frac{d_s}{d}$ and taking the NS5-brane
 limit \eqref{DSLgrav}, we obtain
\begin{align}
 S_{es}
 &=-\frac{1}{4} \int_{-\infty}^{\infty}dz  \int_{0}^{\infty}rdr
 \Bigg[
 -\sum_{s,n}\sum_{t,m} V_{s,n}(r,z)\bigtriangleup V_{t,m}(r,z)
 \nonumber \\
 &\qquad
 -2\sum_{s,n}\left( 
 \frac{1}{g_0}\sin\frac{\pi z}{d}I_0(\frac{\pi r}{d})
 -\tilde C_s\frac{z}{|z|}
 \right) \bigtriangleup V_{s,n}(r,z)
 \Bigg]
 -2\sum_{s=1}^{\Lambda}\tilde C_s Q_{s},
\end{align}
up to irrelevant constants. 
One can see that this gives the action of the electrostatic potential for LST.
In terms of the charge densities $f_s(u)$, it is rewritten as
\begin{align}
 S_{es}&=
 -\frac{2}{\pi}\Bigg[
 \sum_{s=1}^{\Lambda}\sum_{n=-\infty}^{\infty}\frac{1}{2}
 \int du\, f_{s}(u)^2
 \nonumber \\
 &\qquad\qquad
 -\sum_{s,n}\sum_{t,n'}\frac{1}{2\pi} \int du du' \bigg[
 \frac{|d_s+d_t+2n'd|}{(d_s+d_t+2n'd)^2+(u-u')^2}
 \nonumber \\
 &\qquad\qquad\qquad\qquad\qquad\qquad\qquad
 -\frac{|d_s-d_t+2n'd|}{(d_s-d_t+2n'd)^2+(u-u')^2}\bigg] f_{s}(u)f_{t}(u') 
 \nonumber \\
 &\qquad\qquad
 +\frac{1}{g_0}\sum_{s,n}\sin\frac{\pi d_s}{d}\int du\, \cosh\frac{\pi u}{d} f_{s}(u)
 -\sum_{s=1}^{\Lambda}\tilde C_s\left(\int du\, f_{s}(u)-\pi Q_{s}\right)
 \Bigg].
\end{align}
It is easily seen that the equation of motion for $f_{s}(u)$ coincides
with the integral equation of $\tilde f_s(u)$ in \eqref{tildefiIntEq}. 
Thus, in the limit \eqref{DSLgrav}, 
the action of the electrostatic system for PWMM indeed 
reduces to the action for LST. 
%Hence, in the NS5 limit, $f_s(u)=\tilde f_s(u)$ follows.

%%%%%%%%%%%%%%%%%%%%%%%%%%%%%%%%%%%%%%%%%%%%%%%%%%%%%
\section{Details of numerical analysis}

In this appendix, we show our choice of the parameters
for the numerical computation in section {\ref{Numerical analysis}}.

\subsection{Parameter region for the NS5-brane limit}
\label{sec:parameter_region for the NS5-brane limit}

The expectation values are computed with
\beq
	N^{(2)}_2=20,40,60,80,100
\eeq 
at each fixed value of $\tilde g_s=100,200,300,400,500$.
We choose the other fixed parameters as
% $N^{(1)}_5=1$, $N^{(2)}_5=2$ 
$D_1=N^{(1)}_5=1$, $D_2=D=2$ 
and $N^{(1)}_2=4$.
These are chosen so that $\lambda\gg D^{4}$ in \eqref{limit},
which is actually required by the NS5-brane limit \eqref{DSLgauge}
% (\ref{eq:NS5-brane_limit_numerical})
as $\lambda$ scales as $(D \ln N_2^{(2)})^4$
and hence $\lambda \gg D^4$. 

Note, however, that $D_1$ and $D_2-D_1$ are not large but 1 in this case.
Thus we expect that the system reproduces the theory on two NS5-branes
with a D2-brane flux,
which should contain substantial string and quantum correction.
This is not a problem because the purpose of the simulation is
to confirm the double scaling limit.%HERE

\subsection{Parameter region for the large-$N$ limit}
\label{sec:parameter_region for the large-N limit}

We compute the expectation values by setting the parameters $N_2^{(2)}$ and $\lambda = g^2 N_2^{(2)}$
to the values listed in Tab.~\ref{tab:gs_200-500} and \ref{tab:gs_600-900}.
These tables show the corresponding $\tilde g_s$ obtained by \eqref{DSLgauge}.

We choose the other fixed parameters as
% $N^{(1)}_5=3$, $N^{(2)}_5=D=6$
$D_1=N^{(1)}_5=3$, $D_2=D=6$ 
and $N^{(1)}_2=4$.
Again, $\lambda$ needs to satisfy $\lambda\gg D^{4}$.
Since it is a stringent condition
for numerical simulations %, at least, in our machine resource 
due to costly computation of such large-size matrices,
we choose the region of $\lambda$ to be
% $\lambda > (N^{(2)}_5)^{4}/8 = 162$.
$\lambda > D^{4}/8 = 162$.

\begin{table}[htbp]
\centering
\caption{\small
 $N_2^{(2)}$ and $\lambda$ ($= 200, 300, 400, $ and 500) which we choose in our simulations. The cell whose row and column are $\lambda$ and $N_2^{(2)}$, respectively, indicates the corresponding value of $\tilde g_s$ computed through \eqref{DSLgauge}.
}
\begin{tabular}{|c||c|c|c|c|c|c|c|c|c|c|c|}\hline
	{$\lambda$}\textbackslash {$N_2$}
	&70		&80		&90		&100	&110	&120	&130	&140	&150	&200	&300	\\ \hline\hline
	200	&10.75	&9.40	&8.36	&7.52	&6.84	&6.27	&5.79	&5.37	&5.02	&3.76	&2.51	
	\\ \hline 	
	300	&19.71	&17.25	&15.33	&13.80	&12.54	&11.50	&10.61	&9.86	&9.20	&6.90	&4.60	
	\\ \hline 
	400	&31.01	&27.13	&24.12	&21.71	&19.73	&18.09	&16.70	&15.50	&14.47	&10.85	&7.24	
	\\ \hline
	500	&44.69	&39.10	&34.76	&31.28	&28.44	&26.07	&24.06	&22.34	&20.85	&15.64	&10.43	
	\\ \hline
\end{tabular}
\label{tab:gs_200-500}
\end{table}

\begin{table}[htbp]
\centering
\caption{\small
$N_2^{(2)}$ and $\lambda$ ($= 600, 700, 800, $ and 900) which we choose in our simulations. Again, each cell in the body of the table indicates the corresponding value of $\tilde g_s$ computed through \eqref{DSLgauge}.
% The values of $N_2^{(2)}, \lambda,$ and corresponding $\tilde g_s$. ($\lambda = 600, 700, 800, $ and 900)
}
\begin{tabular}{|c||c|c|c|c|c|c|c|c|c|}\hline
	{$\lambda$}\textbackslash {$N_2$}
		&70		&90		&110	&130	&150	&200	&300 	&400	&500	\\ \hline\hline
	600	&60.81	&47.30	&38.70	&32.75	&28.38	&21.29	&14.19	&--	&--
	\\ \hline
	700	&79.47	&61.81	&50.57	&42.79	&37.09	&27.81	&18.54	&--	&--
	\\ \hline
	800	&100.75	&78.36	&64.11	&54.25	&47.01	&35.26	&23.51	&17.63	&--
	\\ \hline
	900	&--	&97.01	&79.37	&67.16	&58.21	&43.66	&29.10	&21.83	&17.46
	\\ \hline 	
\end{tabular}
\label{tab:gs_600-900}
\end{table}

%%%%%%%%%%%%%%%%%%%%%%%%%%%%%%%%%%%%%%%%%%%%%%%%%%%%%
\subsection{Coefficients $d_0(\lambda)$ and $d_1(\lambda)$}
\label{coefficients d0 d1}

Here, we show the obtained data of $d_0(\lambda)$ and $d_1(\lambda)$
for $\langle\Tr M^{\; 2}_1\rangle_M$.
% $\langle\Tr M^4_1\rangle$
% and $\langle\Tr e^{-M_1}\rangle$.
% For each VEV,  
We perform large-$N_2$ extrapolation, namely, fitting by a quadratic function $a + b/N_2 + c/(N_2)^2$ where $a, b, c$ are free parameters and $N_2 = N_2^{(2)}$.
As a result of the fitting, we identify $a, b$ with $d_0(\lambda)$ and $d_1(\lambda)$, respectively.

\begin{table}[htp]
\begin{center}
\caption{\small The coefficients of $\langle\Tr M^{\;2}_1\rangle_M$.}
\label{coefficients of TrM^2 ds}
	\begin{tabular}{ccc}\hline
	$\lambda$&$d_0(\lambda)$&$d_1(\lambda)$ \\
	\hline
	$200$ & $0.17(1)$ & $0.236(5)$ \\
	$300$ & $0.28(3)$ & $0.193(5)$ \\
	$400$ & $0.36(5)$ & $0.170(5)$ \\
	$500$ & $0.64(6)$ & $0.139(5)$ \\
	$600$ & $0.73(8)$ & $0.130(5)$ \\
	$700$ & $0.82(11)$ & $0.139(5)$ \\
	$800$ & $0.84(8)$ & $0.115(3)$ \\
	$900$ & $0.93(8)$ & $0.110(3)$ \\
	\hline
	\end{tabular}
\end{center}
\end{table}

%\begin{table}[htp]
%\begin{center}
%\label{coefficients of TrM^2}
%\caption{The coefficients of $\langle\Tr M^4_2\rangle$.}
%	\begin{tabular}{ccc}\hline
%	$\lambda$&$d_0(\lambda)$&$d_1(\lambda)$ \\
%	\hline
%	$200$ & $0.23(1)$ & $2.3(4)\times10^{4}$ \\
%	$300$ & $0.14(4)$ & $6.6(1)\times10^{4}$ \\
%	$400$ & $0.35(8)$ & $1.2(3)\times10^{5}$ \\
%	$500$ & $1.2(1)$ & $1.7(8)\times10^{5}$ \\
%	\hline
%	\end{tabular}
%	
%\end{center}
%\end{table}

%\begin{table}[htp]
%\begin{center}
%\label{coefficients of TrM^2}
%\caption{The coefficients of $\langle\Tr e^{-M_2}\rangle$.}
%	\begin{tabular}{ccc}\hline
%	$\lambda$&$d_0(\lambda)$&$d_1(\lambda)$ \\
%	\hline
%	$200$ & $3.3(1)$ & $8.8(4)\times10^{3}$ \\
%	$300$ & $3.6(2)$ & $1.3(7)\times10^{4}$ \\
%	$400$ & $3.5(3)$ & $2.8(1)\times10^{4}$ \\
%	$500$ & $3.8(9)$ & $3.1(4)\times10^{4}$ \\
%	\hline
%	\end{tabular}
%\end{center}
%\end{table}

%%%%%%%%%%%%%%%%%%%%%%%%%%%%%%%%%%%%%%%%%%%%%%%%%%%%%
\subsection{Comparison with $\tilde g_s$}
\label{comparing_with_gs}

We perform the large-$\lambda$ extrapolation varying the range of $\tilde g_s$
and compare the dependence of the range.

Below $\lambda = 500$, the restriction for $\tilde g_s$ provides no condition.
For $\lambda = 600$, the restriction gives the same condition, and hence, 
the same coefficients for the fit.
The error tends bigger than the one without restriction due to the reduction of the data points for the fit.

\begin{table}[htp]
\begin{center}
\caption{\small The coefficients of $\langle\Tr M^{\; 2}_1\rangle_M$.}
\label{coefficients of TrM^2 g_s}
	\begin{tabular}{c|cc|cc}\hline
	&\multicolumn{2}{c|}{$\tilde g_s \le 50$} & \multicolumn{2}{c}{$\tilde g_s \le 60$}
	\\ \hline
	$\lambda$ &$d_0(\lambda)$&$d_1(\lambda)$ & $d_0(\lambda)$&$d_1(\lambda)$ \\
	\hline
	$600$ & $0.63(1)$ & $0.139(8)$ & $0.63(1)$ & $0.139(8)$ \\
	$700$ & $0.5(3)$ & $0.14(2)$ & $0.4(2)$ & $0.15(1)$ \\
	$800$ & $0.7(2)$ & $0.13(2)$ & $0.7(2)$ & $0.13(1)$ \\
	$900$ & $1.1(3)$ & $0.10(2)$ & $1.1(3)$ & $0.10(2)$ \\
	\hline
	\end{tabular}
\end{center}
\end{table}

\begin{figure}[htpb]
	% \begin{minipage}{0.475\textwidth}
	% \centering
	% \includegraphics[width=\textwidth]{ratio_trM2_mod_vs_lambda_gs_le_50.pdf}
	% \end{minipage}
	\begin{minipage}{0.475\textwidth }
	\centering
	\includegraphics[width=\textwidth]{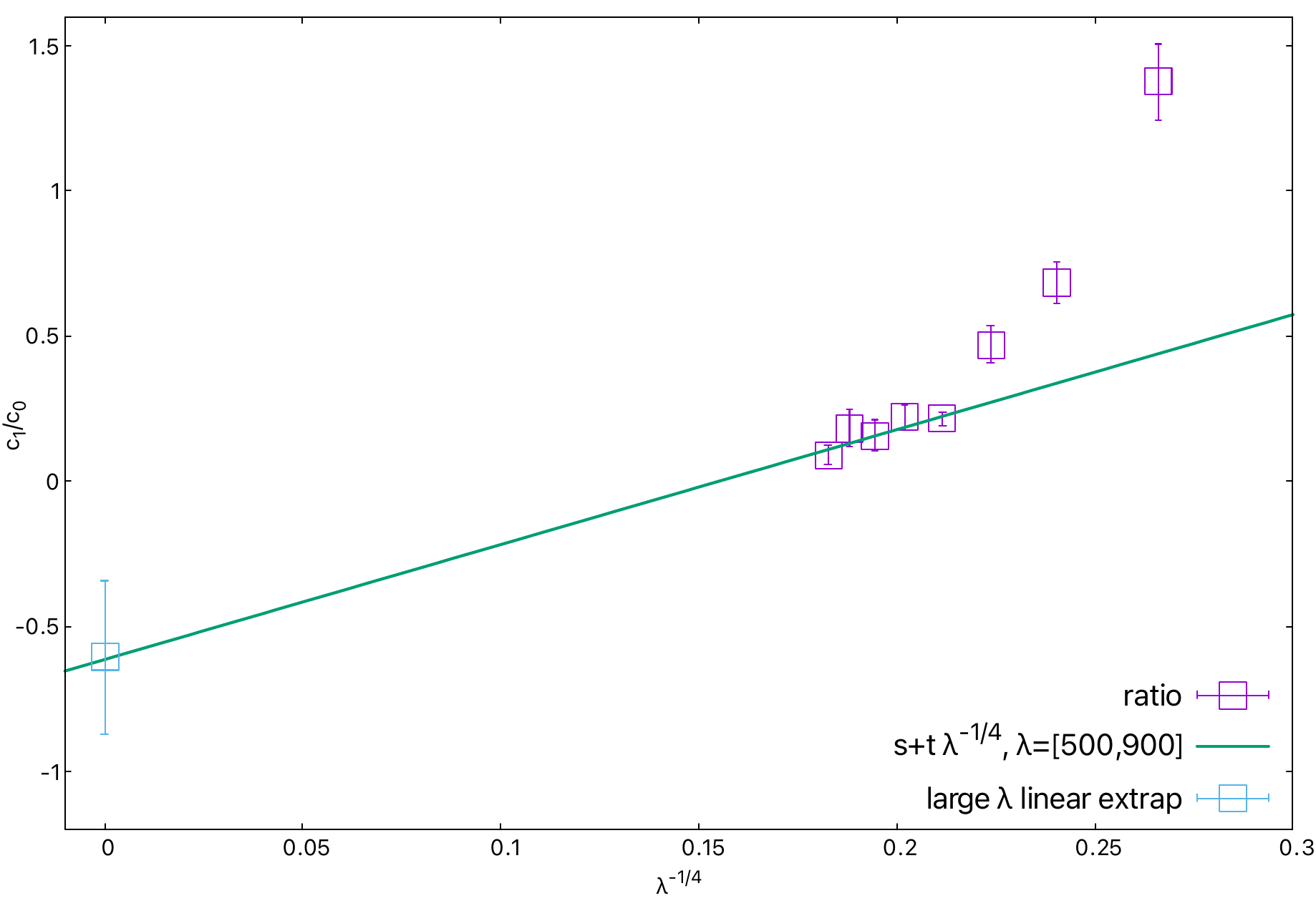}
	\\
	(a) $\tilde g_s \le 50$
	\end{minipage}
	% \\
	% \centering
	% (a) $\tilde g_s \le 50$
	% \vspace{20pt}
	%
	% \begin{minipage}{0.475\textwidth}
	% \centering
	% \includegraphics[width=\textwidth]{ratio_trM2_mod_vs_lambda_gs_le_60.pdf}
	% \end{minipage}
	\begin{minipage}{0.475\textwidth }
	\centering
	\includegraphics[width=\textwidth]{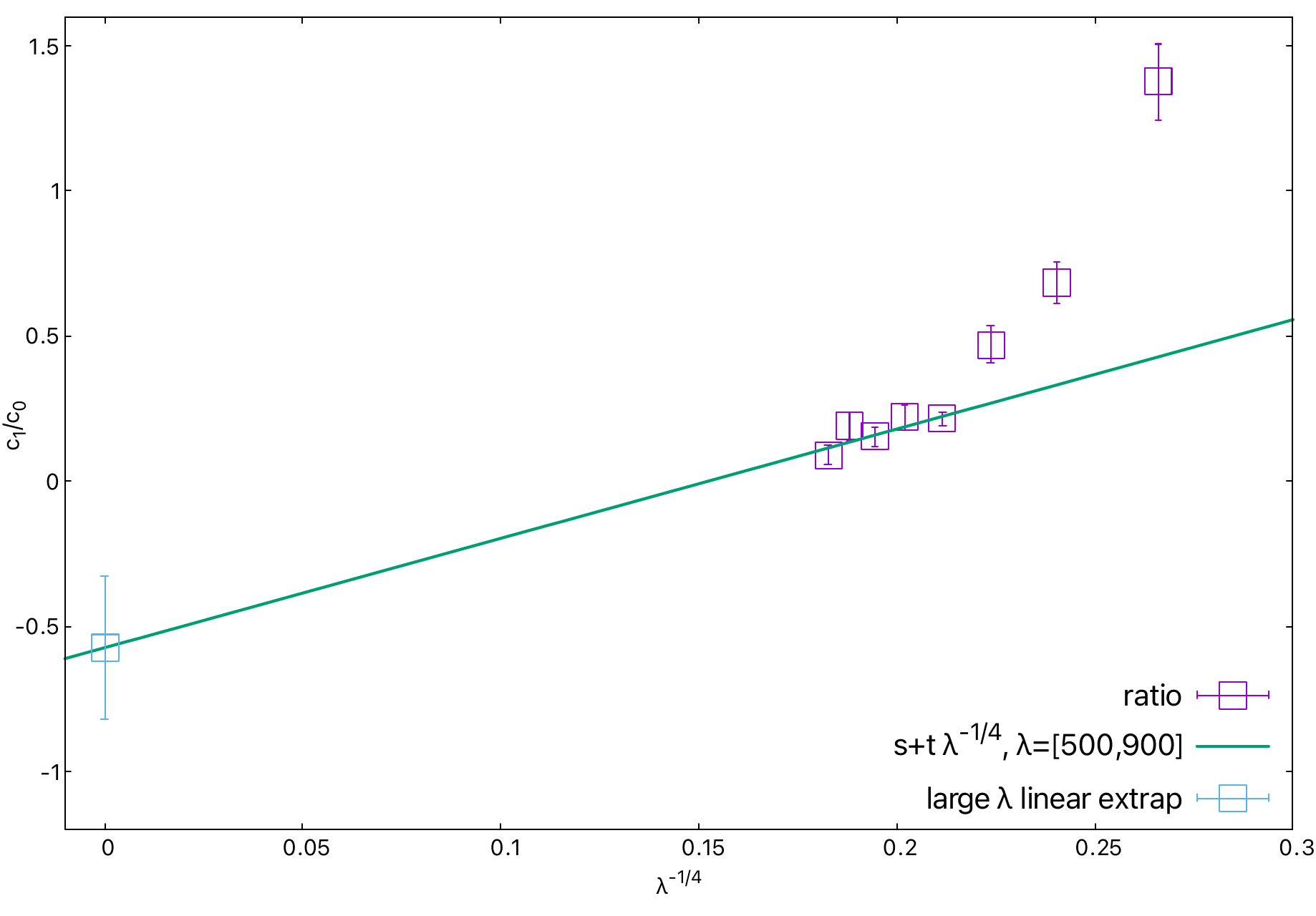}
	\\
	(b) $\tilde g_s \le 60$
	\end{minipage}
	% \\
	% \centering
	% (b) $\tilde g_s \le 60$
	
	\caption
	{\small The plot of the ratio $c_1/c_0$ of
	$\langle\Tr M^{\;2}_1\rangle_M$.
	At each data point of $\lambda$, we restrict $\tilde g_s \le 50$ in (a) and $\tilde g_s \le 60$ in (b) for the extrapolation.
        The horizontal axis is $\lambda^{-1/4}$.
	The green lines are the fitted lines %for different regions of $\lambda$, 
	by $s + t \lambda^{-1/4}$,
        % and $s + t \lambda^{-1/4} + u \lambda^{-1/2}$, respectively
        where $s$ and $t$ are fitting parameters.
	% [Left] The horizontal axis is $\lambda$. 
	% [Right] 
	The blue square at $\lambda^{-1/4} = 0$ shows the extrapolated value of $c_1/c_0$ by the 't~Hooft limit.
	}
	\label{fig:ratio_trM2_vs_lambda_gs_le_50-60}
\end{figure}


\begin{thebibliography}{99}
\bibitem{Maldacena}
J.~M.~Maldacena, 
%``The large N limit of superconformal field theories and supergravity,''
Adv.\ Theor.\ Math.\ Phys.\  {\bf 2} (1998) 231. 
%[arXiv:hep-th/9711200]. 

\bibitem{GKP}
S.~S.~Gubser, I.~R.~Klebanov and A.~M.~Polyakov,
%``Gauge theory correlators from non-critical string theory,''
Phys.\ Lett.\ B {\bf 428} (1998) 105. 
%[arXiv:hep-th/9802109]. 

\bibitem{Witten}
E.~Witten, 
%``Anti-de Sitter space and holography,''
Adv.\ Theor.\ Math.\ Phys.\  {\bf 2} (1998) 253. 
%[arXiv:hep-th/9802150].

\bibitem{LSTreview}
% \bibitem{Aharony:1999ks} 
  O.~Aharony,
  %``A Brief review of 'little string theories',''
  Class.\ Quant.\ Grav.\  {\bf 17}, 929 (2000)
  [hep-th/9911147];
  %%CITATION = HEP-TH/9911147;%%
  %111 citations counted in INSPIRE as of 29 Apr 2014
% \bibitem{Kutasov:2001uf} 
  D.~Kutasov,
  ``Introduction to little string theory,''
  Superstrings and related matters. Proceedings, Spring School, Trieste, Italy, April 2-10, 2001;
  %%CITATION = INSPIRE-572494;%%
  and references therein.

%\cite{Berenstein:2002jq}
\bibitem{Berenstein:2002jq} 
  D.~E.~Berenstein, J.~M.~Maldacena and H.~S.~Nastase,
  %``Strings in flat space and pp waves from N=4 superYang-Mills,''
  JHEP {\bf 0204}, 013 (2002).
%  doi:10.1088/1126-6708/2002/04/013
%  [hep-th/0202021].
  %%CITATION = doi:10.1088/1126-6708/2002/04/013;%%
  %1586 citations counted in INSPIRE as of 03 Aug 2016

\bibitem{LM} 
  H.~Lin and J.~M.~Maldacena,
  %``Fivebranes from gauge theory,''
  Phys.\ Rev.\ D {\bf 74}, 084014 (2006)
  [hep-th/0509235].
  %%CITATION = HEP-TH/0509235;%%

\bibitem{Ling:2006up} 
  H.~Ling, A.~R.~Mohazab, H.~-H.~Shieh, G.~van Anders and M.~Van Raamsdonk,
  %``Little string theory from a double-scaled matrix model,''
  JHEP {\bf 0610}, 018 (2006)
  [hep-th/0606014].
  %%CITATION = HEP-TH/0606014;%%
  %13 citations counted in INSPIRE as of 21 Jun 2013

\bibitem{LLM} 
  H.~Lin, O.~Lunin and J.~M.~Maldacena,
  %``Bubbling AdS space and 1/2 BPS geometries,''
  JHEP {\bf 0410}, 025 (2004)
  [hep-th/0409174].
  %%CITATION = HEP-TH/0409174;%%
  %463 citations counted in INSPIRE as of 01 Dec 2013

\bibitem{Ishiki:2006yr}
  G.~Ishiki, S.~Shimasaki, Y.~Takayama and A.~Tsuchiya,
  %``Embedding of theories with SU(2|4) symmetry into the plane wave matrix
  %model,''
  JHEP {\bf 0611} (2006) 089.
%  [arXiv:hep-th/0610038].
  %%CITATION = JHEPA,0611,089;%%

 \bibitem{Ishiki:2006rt} 
   G.~Ishiki, Y.~Takayama and A.~Tsuchiya,
   %``N=4 SYM on R x S**3 and theories with 16 supercharges,''
   JHEP {\bf 0610}, 007 (2006).
%%   [hep-th/0605163].
%%   %%CITATION = HEP-TH/0605163;%%
%%   %33 citations counted in INSPIRE as of 20 Dec 2013

\bibitem{Ling:2006xi} 
  H.~Ling, H.~-H.~Shieh and G.~van Anders,
  %``Little String Theory from Double-Scaling Limits of Field Theories,''
  JHEP {\bf 0702}, 031 (2007).
%  [hep-th/0611019].
  %%CITATION = HEP-TH/0611019;%%
  %6 citations counted in INSPIRE as of 08 Jul 2013

\bibitem{Asano:2014vba} 
  Y.~Asano, G.~Ishiki, T.~Okada and S.~Shimasaki,
  %``Emergent bubbling geometries in the plane wave matrix model,''
  JHEP {\bf 1405}, 075 (2014).
%  [arXiv:1401.5079 [hep-th]].
  %%CITATION = ARXIV:1401.5079;%%

\bibitem{Asano:2014eca} 
  Y.~Asano, G.~Ishiki and S.~Shimasaki,
  %``Emergent bubbling geometries in gauge theories with SU(2|4) symmetry,''
  JHEP {\bf 1409}, 137 (2014).
%  [arXiv:1406.1337 [hep-th]].
  %%CITATION = ARXIV:1406.1337;%%
  %1 citations counted in INSPIRE as of 21 Apr 2015

\bibitem{vanAnders:2007ky} 
  G.~van Anders,
  %``General Lin-Maldacena solutions and PWMM Instantons from supergravity,''
  JHEP {\bf 0703}, 028 (2007)
  [hep-th/0701277 [HEP-TH]].
  %%CITATION = HEP-TH/0701277;%%
  %8 citations counted in INSPIRE as of 22 May 2015

\bibitem{Maldacena:2002rb}
  J.~Maldacena, M.~M.~Sheikh-Jabbari and M.~Van Raamsdonk,
  %``Transverse fivebranes in matrix theory,''
  JHEP {\bf 0301} (2003) 038
  [arXiv:hep-th/0211139].

\bibitem{Pestun:2007rz} 
  V.~Pestun,
  %``Localization of gauge theory on a four-sphere and supersymmetric Wilson loops,''
  Commun.\ Math.\ Phys.\  {\bf 313}, 71 (2012).
%  [arXiv:0712.2824 [hep-th]].
  %%CITATION = ARXIV:0712.2824;%%
  %332 citations counted in INSPIRE as of 01 Dec 2013

\bibitem{Asano:2012zt} 
  Y.~Asano, G.~Ishiki, T.~Okada and S.~Shimasaki,
  %``Exact results for perturbative partition functions of theories with SU(2|4) symmetry,''
  JHEP {\bf 1302}, 148 (2013).
%  [arXiv:1211.0364 [hep-th]].
  %%CITATION = ARXIV:1211.0364;%%
  %1 citations counted in INSPIRE as of 21 Jun 2013

\bibitem{Asano:2017xiy} 
  Y.~Asano, G.~Ishiki, S.~Shimasaki and S.~Terashima,
  % ``On the transverse M5-branes in matrix theory,''
  Phys. Rev. D {\bf 96}, 126003 (2017).
  % [arXiv:1701.07140 [hep-th]].

\bibitem{Asano:2017nxw}  
  Y.~Asano, G.~Ishiki, S.~Shimasaki and S.~Terashima,
  % ``Spherical transverse M5-branes from the plane wave matrix model,''
  JHEP {\bf 1802}, 076 (2018).
  % [arXiv:1711.07681 [hep-th]].

\bibitem{Roychowdhury:2021unp}
  D.~Roychowdhury,
  %``Wilson loops for non-Abelian T duality and matrix models,''
  [arXiv:2110.05395 [hep-th]].


\end{thebibliography}
\end{document}